\documentclass[apj]{emulateapj}
\usepackage{bm}
\newcommand{\bjdtdb}{${\rm {BJD_{TDB}}}$}
\newcommand{\feh}{{\left[{\rm Fe}/{\rm H}\right]}}

\newcommand{\msun}{${\rm M}_\Sun$}
\newcommand{\rsun}{${\rm R}_\Sun$}

\newcommand{\mj}{${\,{\rm M}_{\rm J}}$}
\newcommand{\rj}{${\,{\rm R}_{\rm J}}$}

\newcommand{\three}{3.6$\mu$m\ }
\newcommand{\four}{4.5$\mu$m\ }
\newcommand{\um}{$\mu$m\ }
\newcommand{\water}{$\mathrm{H}_2\mathrm{O}$\ }

\begin{document}

\title{Evidence for Atmospheric Cold-trap Processes in the Noninverted Emission Spectrum of Kepler-13Ab Using HST/WFC3}

\author{Thomas G.\ Beatty\altaffilmark{1,2}, Nikku Madhusudhan\altaffilmark{3}, Angelos Tsiaras\altaffilmark{4}, Ming Zhao\altaffilmark{1,2}, Ronald L. Gilliland\altaffilmark{1,2}, Heather A. Knutson\altaffilmark{5}, Avi Shporer\altaffilmark{5,6,7}, Jason T. Wright\altaffilmark{1,2}}

\altaffiltext{1}{Department of Astronomy \& Astrophysics, The Pennsylvania State University, 525 Davey Lab, University Park, PA 16802; tbeatty@psu.edu}
\altaffiltext{2}{Center for Exoplanets and Habitable Worlds, The Pennsylvania State University, 525 Davey Lab, University Park, PA 16802}
\altaffiltext{3}{Institute of Astronomy, University of Cambridge, Madingley Road, Cambridge CB3 0HA, UK}
\altaffiltext{4}{Department of Physics \& Astronomy, University College London, Gower Street, WC1E6BT London, UK}
\altaffiltext{5}{Division of Geological and Planetary Sciences, California Institute of Technology, Pasadena, CA 91125, USA}
\altaffiltext{6}{Jet Propulsion Laboratory, California Institute of Technology, 4800 Oak Grove Drive, Pasadena, CA 91109, USA}
\altaffiltext{7}{NASA Sagan Fellow}

\shorttitle{HST/WFC3 Kepler-13Ab Eclipse Observations}
\shortauthors{Beatty et al.}

\begin{abstract}
We observed two eclipses of the Kepler-13A planetary system, on UT 2014 April 28 and UT 2014 October 13, in the near-infrared using Wide Field Camera 3 on the Hubble Space Telescope. By using the nearby binary stars Kepler-13BC as a reference, we were able to create a differential light curve for Kepler-13A that had little of the systematics typically present in HST/WFC3 spectrophotometry. We measure a broadband (1.1\um to 1.65$\mu$m) eclipse depth of $734\pm28$\,ppm, and are able to measure the emission spectrum of the planet at $R\approx50$ with an average precision of 70\,ppm. We find that Kepler-13Ab possesses a noninverted, monotonically decreasing vertical temperature profile. We exclude an isothermal profile and an inverted profile at more than 3\,$\sigma$. We also find that the dayside emission of Kepler-13Ab appears generally similar to an isolated M7 brown dwarf at a similar effective temperature. Due to the relatively high mass and surface gravity of Kepler-13Ab, we suggest that the apparent lack of an inversion is due to cold-trap processes in the planet's atmosphere. Using a toy model for where cold-traps should inhibit inversions, and observations of other planets in this temperature range with measured emission spectra, we argue that with more detailed modeling and more observations we may be able to place useful constraints on the size of condensates on the daysides of hot Jupiters.  
\end{abstract}

\section{Introduction}

The most immediately measurable property of an exoplanet's emission is its temperature. Both broadband and spectroscopic observations are fundamentally measuring the brightness temperature of the atmosphere at a given wavelength, whether directly if the planet is actually imaged, or more commonly relative to the temperature of its host star when a planet goes into eclipse. How we relate the observed brightness temperature, as a function of wavelength, to the physical properties of an exoplanet's atmosphere is the crux of atmospheric characterization. 

Unlike the majority of stars and brown dwarfs, the transformation between observed brightness temperature and physical properties for exoplanets is complicated by the fact that the dominant energy source in the atmosphere is usually external: irradiation from a host star. Thus, while with a stellar spectrum one can usually assume that hotter temperatures imply light at those wavelengths emerges from deeper within the stellar atmosphere, for exoplanets this is not the case. For example, Earth and the giant Solar System planets all possess some sort of temperature inversion in their atmospheres, where the temperature begins to increase with height. For Earth, this causes sharp emission features in the centers of the 9.5\um O$_3$ and the 15\um CO$_2$ absorption bands. Without an understanding of the temperature structure of the Earth's atmosphere, both these features would be difficult to interpret.

Understanding what determines the vertical temperature structure of exoplanets, and hot Jupiters in particular, has thus been one of the major observational and theoretical tasks of the past decade. Based on early observations \citep[e.g.,][]{knutson2008} and expectations based on the Solar System planets, it was initially believed that all hot Jupiters hotter than approximately 1800\,K should possess temperature inversions in their atmospheres \citep[e.g.,][]{fortney2008}.

Generically, a temperature inversion requires a strong optical absorber that also increases the grayness of thermal opacities in the atmosphere and impedes cooling \citep{hubeny2003,fortney2008,parmentier2015}. Observations of field brown dwarfs show clear signatures of gas-phase TiO and VO in the optical -- both of which meet the necessary criteria -- and so these two molecules are believed to be the primary drivers of potential inversions in hot Jupiters' atmospheres \citep{hubeny2003,fortney2008}. At the typical pressures of a hot Jupiter's stratosphere, both TiO and VO become gases at approximately 1800\,K.

However, subsequent observations have revealed no clear evidence for temperature inversions in hot Jupiters with daysides cooler than 2500\,K \citep[e.g.,][]{madhu2014,crossfield2015}. The lack of inversions in hot Jupiters led to suggestions that UV radiation from the host stars may be photodisassociating TiO/VO \citep{knutson2010}, or that TiO/VO may be condensing and settling out of the atmosphere on the dayside \citep{spiegel2009} or the nightside \citep{parmentier2013}. Recently, \cite{wakeford2016a} suggested that gaseous TiO/VO may still be present in hot Jupiters' atmospheres, but obscured by high altitude clouds. Currently, there is no clear consensus to explain the lack of strong inversion signals in the surveyed hot Jupiters.

The observational effort to understand the temperature structure in hot Jupiters' atmosphere has been tremendously aided by the installation of Wide Field Camera 3 (WFC3) on the Hubble Space Telescope (HST). The spectral resolution and precision of the WFC3 observations mean that rather than having to infer the presence of spectral features as is necessary using broadband data, the WFC3 spectra show them. This capability allowed \cite{haynes2015} to present the first clear detection of a temperature inversion in the ultra-hot Jupiter WASP-33b. The high average brightness temperature of WASP-33b's dayside (3300\,K) led them to suggest that TiO/VO driven inversions may only be present in extremely hot giant planets, which is supported by the apparently isothermal WFC3 emission spectra measured for WASP-103b's dayside \citep[2850\,K,][]{cartier2017}, WASP-12b's dayside \citep[2930\,K,][]{swain2013,stevenson2014}, and the inverted dayside of WASP-121b \cite[2700\,K],[]{evans2017}.

To further investigate the vertical temperature structure of ultra-hot Jupiters, we therefore observed two secondary eclipses of Kepler-13Ab \citep{shporer2011} using HST/WFC3. The Kepler-13 system is composed of three stars: the planet host Kepler-13A, and the unresolved binary Kepler-13BC, with the two components separated by $1''.15$ \citep{shporer2014}. Kepler-13A and -13B are both nearly equal mass A-dwarfs, while Kepler-13C is a fainter K-dwarf. The planet Kepler-13Ab has been observed in eclipse before, by \cite{shporer2014}, who measured the broadband emission spectrum in the Kepler bandpass, the $Ks$-band, and the Spitzer \three and \four bands. \cite{shporer2014} measured an average dayside brightness temperature of $2750\pm160$\,K across all four bands, though the results of these observations did not clearly indicate a preferred temperature structure for Kepler-13Ab's dayside.

We were also interested in Kepler-13Ab due to its relatively high mass of $6.52\pm1.58$\mj\ \citep{shporer2014} and correspondingly high surface gravity of 3.3 times that of Jupiter. Recent observations of KELT-1b's emission spectrum by \cite{beatty2016}, which has a dayside temperature of 3150\,K and a surface gravity 22 times Jupiter's, showed a noninverted, monotonically decreasing temperature profile, which led those authors to suggest that surface gravity also plays a strong role in setting the vertical temperature of hot Jupiters. The similar dayside temperature, but lower surface gravity, of Kepler-13Ab therefore presented itself as a relevant comparison object.   

\section{Observations and Data Reduction}

We observed two eclipses of the Kepler-13A (KOI-13, BD+49 2629) planetary system, on UT 2014 April 28 and UT 2014 October 13, using the infrared (IR) mode of WFC3 on the Hubble Space Telescope. The two visits were each composed of five HST orbits. At the beginning of each visit we took a single direct image in the F126N filter to allow us to determine an initial wavelength solution before switching to the G141 grism (1.1\um to 1.7\um) for the remainder of both visits. To decrease the image read-out times we used WFC3's 256$\times$256 pixel subarray mode.

Despite the relative brightness of Kepler-13A ($J=9.466$ and $H=9.455$) we observed both of the eclipses in staring mode, rather than spatial scan mode. This was due to the presence of the nearly equal brightness companion star system Kepler-13BC $1''.15$ \citep{shporer2014}, or approximately 8 pixels, away from the planet-host Kepler-13A. The Kepler-13BC system itself is unresolved in our direct and grism images, and we intended to minimize the blending of Kepler-13A and Kepler-13BC by not scanning during our exposures.

We oriented the detector on the sky such that Kepler-13A and Kepler-13BC were close to being aligned along the detector's pixel columns, perpendicular to the dispersion direction of our spectra. For the April visit, Kepler-13A was above Kepler-13BC, while for the October visit the spacecraft's roll was reversed. We used the SPARS10 and NSAMP=3 readout modes, which gave us an exposure time of 7.6 seconds. The first orbit of each visit took 100 grism exposures, while the subsequent four orbits each collected 101 individual grism exposures, for a total of 504 grism exposures per visit.
  
\subsection{Image Calibration}

We began our image calibration from the \textsc{flt} images provided by \textsc{STScI}. We first flat-fielded the images using the procedure outline in Section 6.2 of the aXe handbook \citep{kummel2011}. This necessitated an initial wavelength solution for our images, which we determined using the method described in Section 2.2, but here using the \textsc{flt} images. We then used this wavelength solution and the flat-field coefficients given in the G141 flat-field data cube provided by \textsc{STScI} to determine, and apply, a flat-field correction over all of our grism images.

We next defined a bad-pixel mask by manually selecting bad pixels on the flat field image, and we corrected these bad pixels on the flat-fielded images by replacing their values with the median values of the eight pixels immediately around the bad pixels. Ultimately, none of the bad pixels lay within our spectroscopic extraction aperture, rendering this step largely unnecessary.

To identify and remove cosmic ray hits, we divided each image into two sections: the area around the stellar spectra, and the surrounding area dominated by the sky background. Within the area around the stellar spectra, which we defined to be 140 pixels wide and 50 pixels tall and centered on the two spectral traces in the first grism image, we began our cosmic ray rejection by creating a model image by median combining the area in all of our flat-fielded and bad-pixel corrected images. We then divided each individual image by this model, and identified cosmic ray hits as anytime a pixel was more than 2.5 times higher than the median-combined model image. We replaced the pixels in the affected images with the value of that pixel in the model image.

For the lower signal-to-noise background-dominated section of the image, we again created a median-combined model image from all of our flat-fielded and bad-pixel corrected images. Instead of dividing each exposure by this model image, here we subtracted it -- after fitting for a scaling factor to account for the changing pedestal value of the background. Taking the standard deviation of each median subtracted exposure, we identified cosmic ray hits as pixels that lay more than four standard deviations above zero. Again, we replaced the pixels in the affected images with the value of that pixel in the model image.

Finally, we calculated and subtracted the background from each of our grism exposures. To do so, we defined two background regions across the bottom (y-pixels 5 to 45) and top (y-pixels 240 to 255) of each of the flat-fielded, bad-pixel and cosmic ray corrected images. Using these two background stripes, we fit for the background in the central portion of each image assuming the background varied as a \textsc{2D} plane, and subtracted this plane fit from each image. We also experimented with estimating the background using the master sky images provided by the Space Telescope Science Institute, and found that our \textsc{2D} plane background estimate was within 5\% of the best fit background estimated using the master images. Since the median background level was 0.52 electrons per second per pixel for the April visit, and 0.40 electrons per second per pixel for the October visit, the difference in the background estimate is negligible relative to the average target counts of approximately 1500 electrons per second per pixel.

This process left us with flat-fielded, bad-pixel and cosmic ray corrected, background subtracted images with which we performed our spectral extraction. 

\subsection{Wavelength Calibration}

We used the direct image taken at the beginning of each of the visits to establish an initial wavelength solution for that visit. To do so, we used the \textsc{daofind} routine implemented in the PhotUtils Python package to determine the x- and y-pixel locations of Kepler-13A and Kepler-13BC on the detector subarray. We then used the wavelength calibration method described by \cite{kuntschner2009} with the adjusted wavelength coefficients determined by \cite{wilkins2014}, to calculate a wavelength solution for each star. In the spectra of both Kepler-13A and Kepler-13BC the Paschen--$\beta$ line at 1.282\um was clearly visible, and we verified the accuracy of our initial wavelength solution using this feature in the first grism exposure of each visit.  

We assumed that this initial wavelength solution was accurate for the first grism exposures in each of the visits, which were taken immediately after the direct image. For each subsequent grism exposure in a visit, we used the spectral alignment method described by \cite{wilkins2014} to estimate a wavelength shift in the spectrum along the dispersion direction. Briefly, \cite{wilkins2014}'s alignment method median combines all of the unshifted out-of-eclipse \textsc{1D} spectra from a visit, and uses this as a reference spectrum against which one measures the wavelength shift of an individual exposure's spectrum. We did an initial extraction of all our \textsc{1D} spectra using the extraction method described in Section 3, and thus computed a master spectrum for each of the two visits. We cross-correlated this master spectrum against each exposure in the visit using a range of shifts from -10 pixels to +10 pixels. We linearly interpolated the master spectrum and fit for a normalization constant at each of the cross-correlation steps. We assumed that the dispersion of the wavelength solution did not change over the course of a visit.

This procedure gave us wavelength shifts for each individual exposure relative to the median combined master spectrum. Based on our assumption that the wavelength solution derived from the direct image is correct for the first grism exposure in a visit, we subtracted the wavelength shift we measured for this first exposure from our cross-correlated master shifts to determine the shift of each individual exposure relative to our initial wavelength solution (Figure \ref{positions}).

\subsection{The Kepler-13 System and Subtracting Kepler-13BC}

As described in the Introduction, both Kepler-13A and Kepler-13B are mid- to late-A dwarfs of nearly equal brightness. \cite{santerne2012} identified a third star in the Kepler-13 system, which they determined to be on a 65.831 day orbit about Kepler-13B and which they estimated contributes about 1\% of the combined system light. 

\begin{deluxetable}{lcc}
\tablecaption{Properties of Kepler-13A,-13B, and -13C}
\tablehead{\colhead{~~~Parameter} & \colhead{Value} & \colhead{Ref.}}
\startdata
\sidehead{Kepler-13A}
\hline
~~~$T_\mathrm{eff}$ (K) & $7650\pm250$ & Sh14\\
~~~$\log(g)$ & $4.2\pm0.5$ & Sh14\\
~~~$\feh$ & $0.2\pm0.2$ & Sh14\\
~~~v$\sin(i)$ (km s$^{-1}$) & $76.96\pm0.61$ & Jo14\\
~~~$M_*$ (\msun) & $1.72\pm0.10$ & Sh14\\
~~~$R_*$ (\rsun) & $1.71\pm0.04$ & Sh14\\
\hline
\sidehead{Kepler-13B}
\hline
~~~$T_\mathrm{eff}$ (K) & $7530\pm250$ & Sh14\\
~~~$\log(g)$ & $4.2\pm0.5$ & Sh14\\
~~~$\feh$ & $0.2\pm0.2$ & Sh14\\
~~~v$\sin(i)$ (km s$^{-1}$) & $63.21\pm1.00$ & Jo14\\
~~~$M_*$ (\msun) & $1.68\pm0.10$ & Sh14\\
~~~$R_*$ (\rsun) & $1.68\pm0.04$ & Sh14\\
\hline
\sidehead{Kepler-13C (orbiting Kepler-13B)}
\hline
~~~$P$ (days) & $65.831\pm0.029$ & Sa12\\
~~~$e$ & $0.52\pm0.02$ & Sa12\\
~~~$M_*$ (\msun) & 0.40\msun -- 0.75\msun & Jo14
\enddata
\tablecomments{Sa12 = \cite{santerne2012}, Sh14 = \cite{shporer2014}, Jo14 = \cite{johnson2014}.}
\label{tab:stellarprops}
\end{deluxetable} 

Initial spectroscopic characterization of Kepler-13A and -13B by \cite{szabo2011} gave estimated effective temperatures for the two stars of 8200\,K to 8500\,K, masses around 2\,\msun, and radii around 2.5\,\rsun. \cite{shporer2014} subsequently collected an independent set of spectral observations, and their analysis gave effective temperatures of around 7500\,K, masses near 1.7\msun, and radii of about 1.7\rsun. Concurrently with \cite{shporer2014}'s measurements, \cite{huber2014} performed a bulk re-analysis of all the Kepler target stars, and found that Kepler-13A was 9100\,K, with a mass of 2.5\,\msun, and a radius of 3\,\rsun.

The lack of agreement on the temperature, mass, and radius of Kepler-13A has complicated previous studies of the planetary system \citep[e.g.,][]{esteves2015}. To distinguish between these three sets of stellar parameters, we compared the stellar density implied by their masses and radii for Kepler-13A to the stellar density measured for Kepler-13A using the Kepler transit light curve \citep{seager2003}. The \cite{szabo2011} properties give a density of 0.17 g cm$^{-3}$ (they gave no uncertainties on their mass and radius estimates), the \cite{huber2014} properties give $0.12\pm0.08$ g cm$^{-3}$, and the \cite{shporer2014} properties give a density of $0.49\pm0.07$ g cm$^{-3}$. The Kepler light curve, as measured by \cite{shporer2014}, gives a density of $0.52\pm0.03$ g cm$^{-3}$. 

\begin{figure}[t]
\vskip 0.00in
\includegraphics[width=1.1\linewidth,clip]{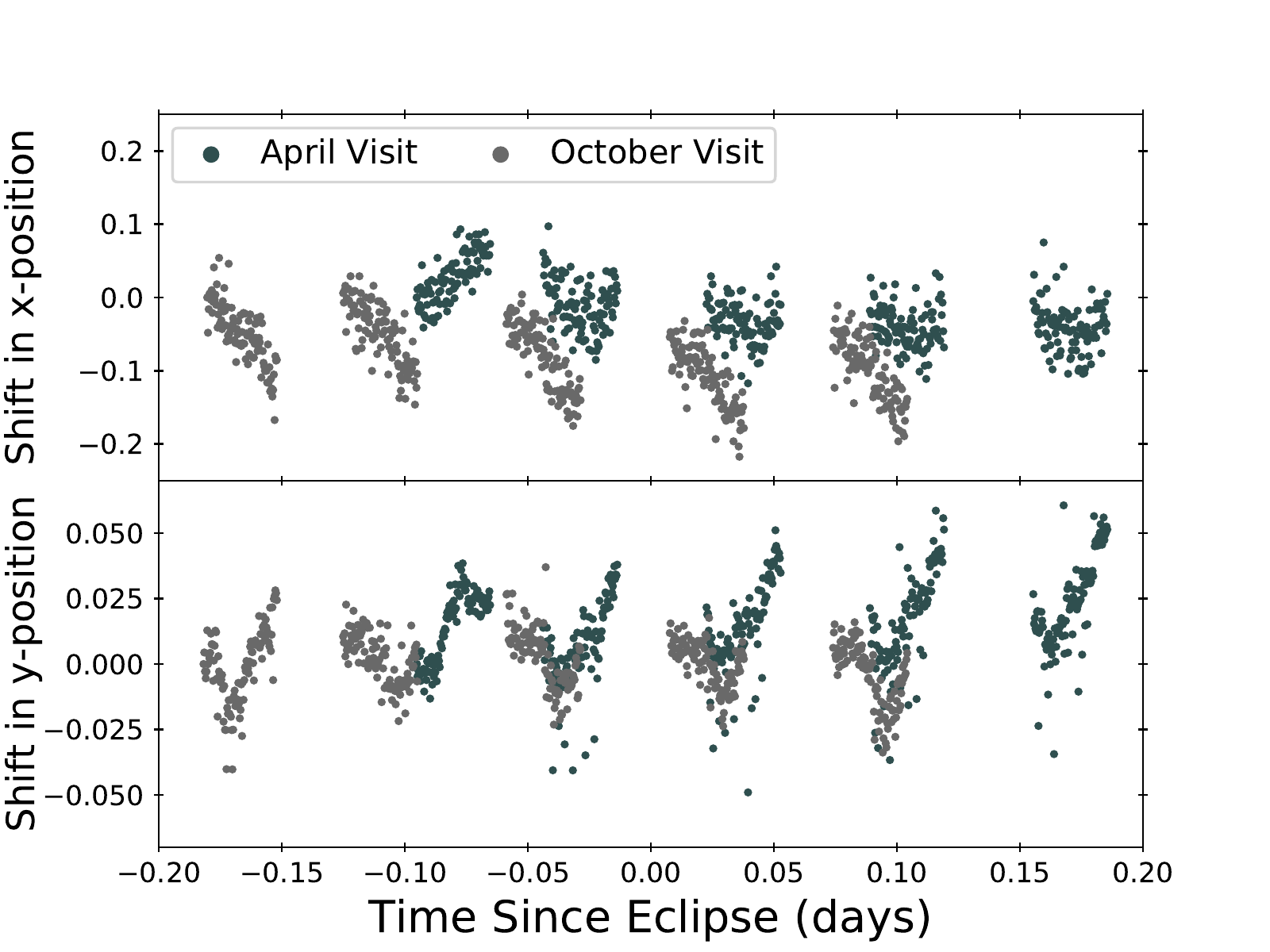}
\vskip -0.0in
\caption{The average x-position (top) and y-position (bottom) of Kepler-13A's spectral trace, relative to the trace's position at the beginning of each visit. The grey-green points show the April visit and the grey points show the October visit. Note that the scale for the panel showing the y-positions is approximately four times smaller than that of the x-position panel.}
\vskip -0.2in
\label{positions}
\end{figure}

Since the \cite{shporer2014} stellar properties are the only set that correctly reproduce the stellar density measured via the transit light curve, we adopted the \cite{shporer2014} stellar properties as correct, and do not consider either the \cite{szabo2011} or \cite{huber2014} estimates in our reduction or analysis. We list the aggregate stellar properties for all three stars in Table \ref{tab:stellarprops}

As mentioned previously, Kepler-13A is separated from Kepler-13BC by $1''.15$ \citep{shporer2014}, or approximately 8 pixels on the WFC3/IR detector. Although the similar spectral types and brightness of Kepler-13A and Kepler-13B led us to expect that there would be relatively little wavelength-dependent dilution in our eclipse observations, we still wished to rigorously account for the effect of the added light from the wings of Kepler-13BC's spectrum.

To do so, we used the \textsc{wayne} simulator \citep{Varley2015} to generate artificial 2D spectra of both stellar components, using the stellar properties listed in Table \ref{tab:stellarprops}. We then subtracted the simulation for one member of the stellar system (e.g., Kepler-13BC) from our observed images to create an undiluted \textsc{2D} spectrum for the remaining member (e.g., Kepler-13A). 

\textsc{wayne} is a simulator for HST/WFC3 spectroscopic images, observed with any of the two infrared grisms (G102 and G141) in both observing modes (staring and spatial scanning). It is able to simulate a number of detector characteristics, such as the read noise, the non-linearity effect, the dark current and the wavelength-dependent flat-field, as well as positioning issues, such as horizontal and vertical shifts and scan speed variations \citep[][and references within]{Varley2015}. A key feature in \textsc{wayne} is the field-dependent structure of the spectrum, as it takes into account the changes in the spectrum trace and the wavelength solution when the spectrum moves on the detector \citep{kuntschner2009}. In addition, to simulate the wavelength-dependent PSF, \textsc{wayne} uses a linear combination of a pair of \textsc{2D} Gaussian distributions, which results in point sources consistent with the PSF ensquared energy fraction given in the Wide Field Camera 3 Instrument Handbook \citep{Dressel2016}.  

For our simulation of the Kepler-13 system, we created two independent simulations, one for each component, and then combined them to a final data set. The position of the simulated spectra was based on the direct (non-dispersed) image of the target and the horizontal and vertical shifts that occurred during the observations (Figure \ref{positions}). In this simulation we included the photon noise, the read noise, the wavelength-dependent flat-field and the sky background, to simulate the \textsc{flt} images. At the stage of the \textsc{flt} images, the remaining reduction steps have already been applied and for this reason we did not include them. In another set of simulations we did not combine the two companions, and also did not include any source of noise or detector characteristics, in order to use them as models for subtracting each companion from the original frames.

As inputs for the simulations we used two synthetic spectra for Kepler-13A and Kepler-13BC, rotationally broadened to the measured $v \sin i$ of the stars (Table \ref{tab:stellarprops}). To create each exposure, the simulator multiplies them by the sensitivity curve of the G141 grism to calculate the expected electron rate as a function of wavelength \citep{Kuntschner2011}. Furthermore, the spectra are scaled based on a combined model for the transit and the systematics \citep[$r_a$, $r_{b1}$, $r_{b2}$ in][Equation 8]{Varley2015}. Finally, the electrons per wavelength channel are distributed on the detector by randomly sampling the wavelength-dependent PSF described above. However, since this scaling is only approximate, in order to subtract the simulated model spectra from the real ones we fit for a scaling factor between them. The parameters used for the models are a combination of the stellar parameters listed in Table \ref{tab:stellarprops} and the planetary parameters listed in Table \ref{tab:broadbandpriors}.

\section{Light Curve Extraction and Fitting}

We extracted light curves for both Kepler-13A and Kepler-13BC. We followed the same procedure for both light curves; for clarity we will describe the process in the context of Kepler-13A's light curve.

For the Kepler-13A light curve, we began with our subtracted images with the \textsc{wayne} simulation of Kepler-13BC's light removed. We fit for the spectral trace of Kepler-13A by fitting a Gaussian profile to the spectrum along detector columns, allowing the baseline level, the profile width, and profile center to all be free parameters. We used the profile centers as the locations of the trace within each column, and fit a 9-degree polynomial to the resultant set of points to determine the y-pixel position of the trace as a function of x-pixel position. Our measured trace positions, and the resulting polynomial fit, are not quite linear, as one would expect from the relations of \cite{kuntschner2009}. However the dispersion in our measured trace about a linear relation (0.085 pixels) is consistent with the dispersion quoted in \cite{kuntschner2009} (0.08 pixels). 

\begin{figure}[t]
\vskip 0.00in
\includegraphics[width=1.1\linewidth,clip]{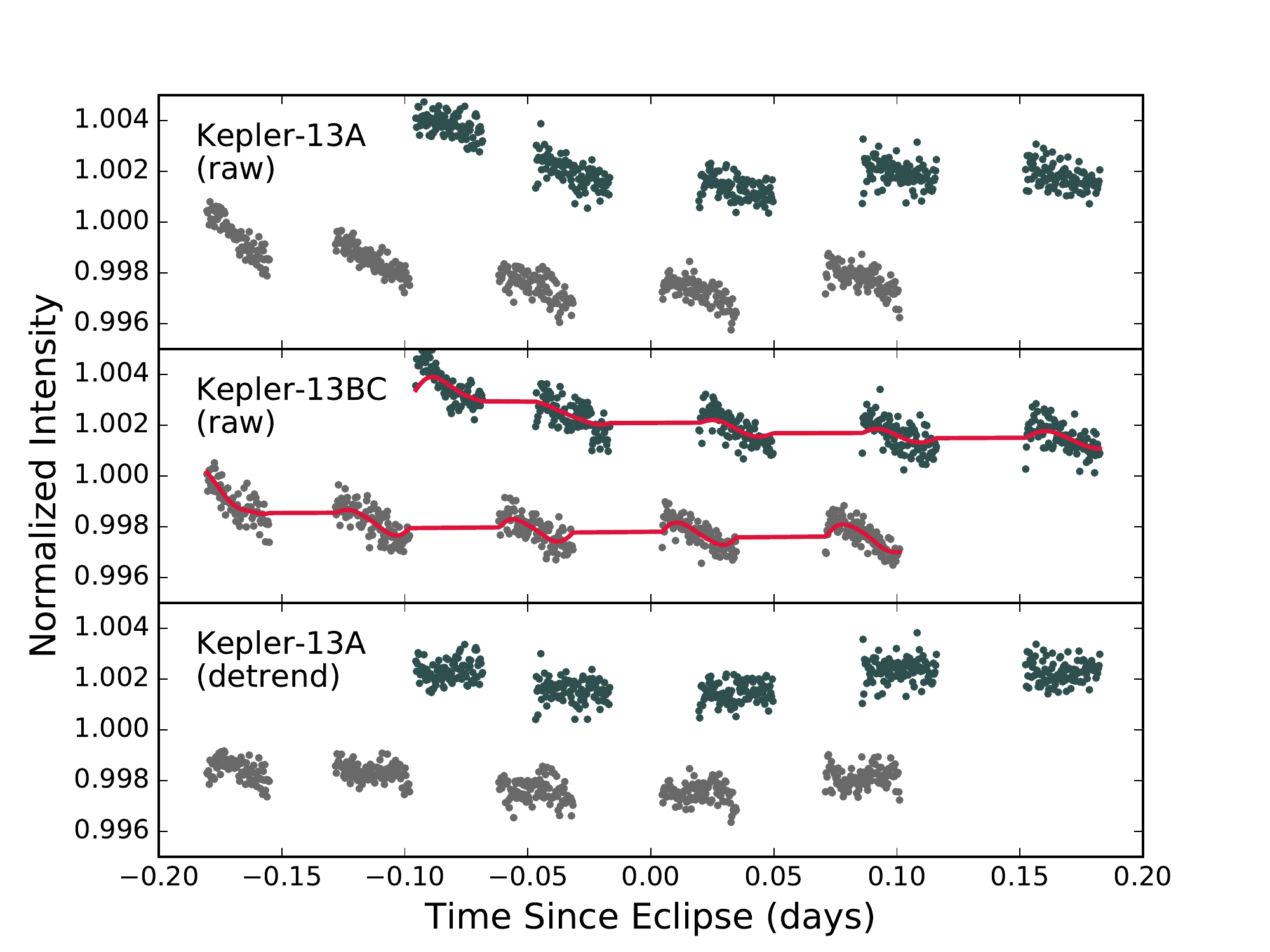}
\vskip -0.0in
\caption{Our raw broadband photometry for Kepler-13A (top panel) on the April (grey-green) and October (grey) visits showed long-term and orbital trends typical for these types of observations. By applying a low-pass filter to the Kepler-13BC photometry (red line, middle panel) and using this as a differential comparison, we were able to create detrended light curves for Kepler-13A that has most of these trends removed (bottom panel).}
\vskip -0.15in
\label{detrendplot}
\end{figure}

We then summed along detector columns to generate a \textsc{1D} spectrum for Kepler-13A using an extraction aperture centered on this polynomial fit to the spectral trace. We used an extraction aperture with a half width of 4.5 pixels about the trace location to sum each of the columns. We chose this aperture size for two reasons. First, though we were confident that our simulated spectra of Kepler-13BC is effectively removing the wings of that system's light, we were concerned that the core of the spectral emission was being imperfectly subtracted. We therefore did not wish to have our extraction aperture for Kepler-13A extend too close to this imperfectly subtracted core. Second, we tested several aperture sizes from 3.0 to 6.0 pixels to determine the optimum aperture size and determine whether the exact choice of aperture significantly affected our final results. We did this by extracting broadband light curves over the range of aperture sizes, and fitting the eclipse using the initial Nelder-Mead likelihood maximization step of our fitting procedure described in Section 3.2. An aperture size of 4.5 pixels results in a fit with the highest likelihood and the lowest scatter in its residuals. We therefore chose 4.5 pixels as the optimum extraction aperture size.

As a test to see how much noise the subtraction process introduced into the \textsc{2D} spectra of Kepler-13A, we examined the wings of the Kepler-13A spectra for asymmetries. To do so, we took the spectral profile in each pixel column, and compared the profile in the top (+2.25 to +4.5 pixels, closer to -13BC) and bottom (-2.25 to -4.5 pixels, farther from -13BC) quarters of our extraction box. Since the outer edge of the top profile is closest to Kepler-13BC, we expected that imperfections in the subtraction process would be most obvious there. We linearly interpolated the counts between individual pixels to determine both the top and bottom profiles at non-integer pixel values, and subtracted the two profiles from each other. Operating under the assumption that the bottom profile was uncontaminated by the subtraction process, we then divided the difference between the profiles by the expected photon-noise in the bottom profile. We then median combined this  ``delta-profile'' over all the extracted columns in each image, and then again over each image for each visit. For both the April and the October visits we find that the median ``delta-profiles'' are at most $1.25\sigma$ away from being perfectly symmetric, with a median deviation of $0.30\sigma$. Since this is below the variation expected from the photon-noise alone, we did not consider the subtraction process to have introduced significant errors into the spectra of Kepler-13A.

With our extracted \textsc{1D} spectra for each image, we used the wavelength solution for each image (Section 2.2) to extract a light curve within the fixed wavelength range of 1.125\um to 1.65$\mu$m. For the broadband data, we summed this entire wavelength range into a single point for each exposure, while for the spectrally-resolved data we subdivided this wavelength range into 15 bins evenly spaced in wavelength.   

\subsection{Differential Fitting via Gaussian Process (GP) Regression}

As usual with HST/WFC3 grism observations, our raw extracted photometry displayed clear correlated noise (top panel of Figure \ref{detrendplot}). As delineated by \cite{wakeford2016}, these are typically a long-term temporal trend over the course of several orbits, an ``L''-shaped hook during the course of an individual orbit, and quick variations within an orbit due to the spacecraft's thermal breathing. Only the first, visit-long slope, and third, thermal breathing, effects were clearly present in our broadband and spectrally-resolved photometry of Kepler-13A and -13BC. 

\begin{figure}[t]
\vskip 0.00in
\includegraphics[width=1.1\linewidth,clip]{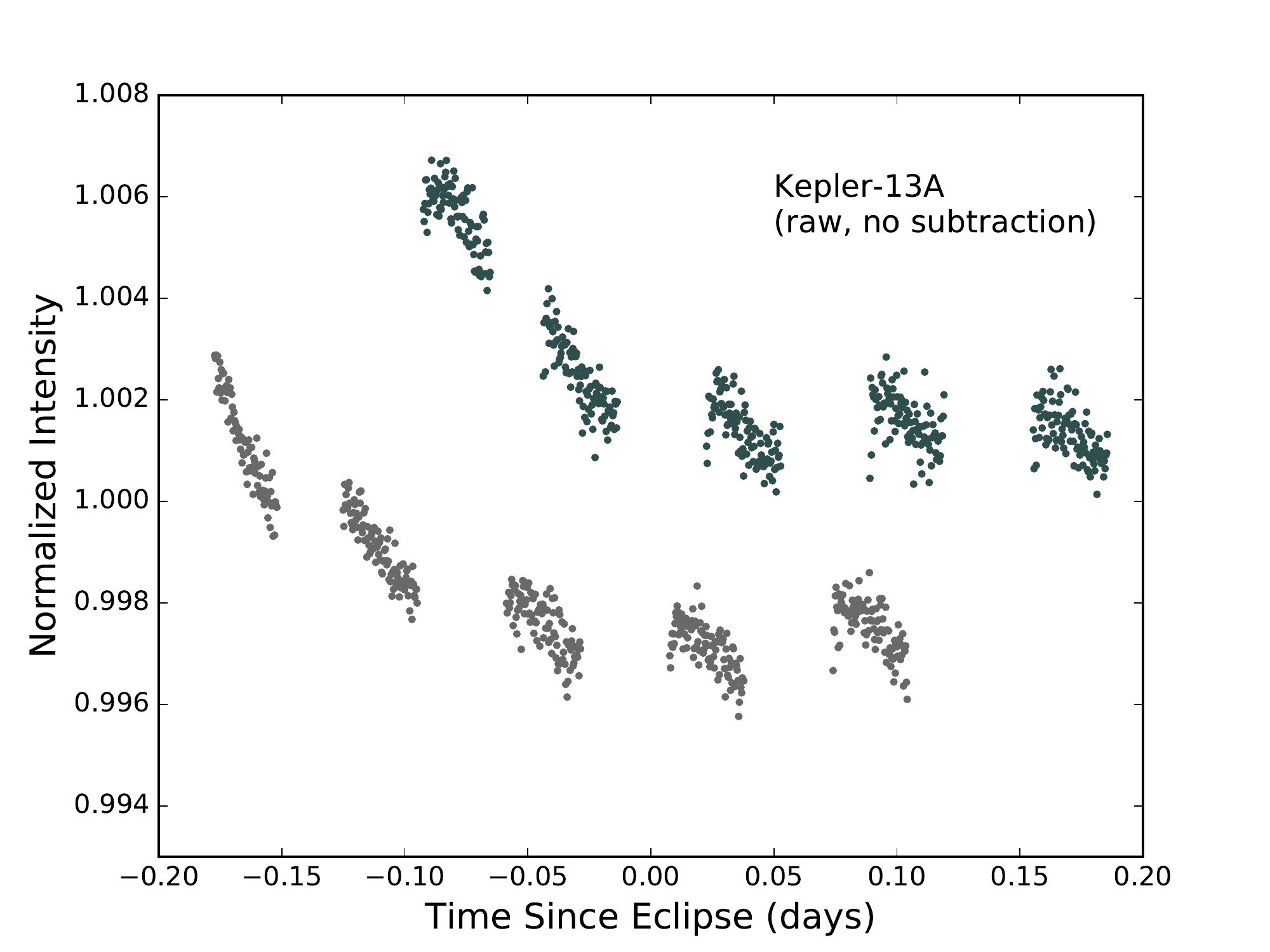}
\vskip -0.0in
\caption{Our raw broadband photometry for Kepler-13A (top panel) on the April (grey-green) and October (grey) visits showed long-term and orbital trends typical for these types of observations. By applying a low-pass filter to the Kepler-13BC photometry (red line, middle panel) and using this as a differential comparison, we were able to create detrended light curves for Kepler-13A that has most of these trends removed (bottom panel).}
\vskip -0.15in
\label{nosub}
\end{figure}

The raw photometry did show a downward trend during the course of each orbit's observations (top panel, Figure \ref{detrendplot}). This is not usually seen in WFC3 observations, but this observation is one of the very few where the SPARS10 sampling sequence at NSAMP=3 was used. In the similar case of WASP-12b \citep{swain2013,mandell2013} a similar behavior can be seen. This may be caused when the detector is continuously flushed to prevent charge build-up after an exposure \citep[][p.\,158]{deustua2016}, in combination with the large number of exposures per orbit. Another possibility is that this trend was introduced by the subtraction process we just described. To test for this, we extracted photometry for Kepler-13A without subtracting off our model for Kepler-13B. As shown in Figure \ref{nosub}, while the no-subtraction photometry shows higher scatter than the subtracted photometry in Figure \ref{detrendplot}, the orbit-long downward trend is still present, and thus not a result of our subtraction process.

At this point the presence of Kepler-13BC greatly simplified our measurement of Kepler-13Ab's eclipse, by providing a nearly equal brightness and equal color comparison light curve with which to perform differential photometry. Since Kepler-13C contributes a negligible amount of light to the -13BC light curve, the broadband response of Kepler-13BC to the HST/WFC3 systematics closely matched the response of Kepler-13A.

We initially attempted to create a differential light curve for Kepler-13A by the straight division of the Kepler-13BC light curve, similar to how we would treat this in ground-based photometry. To our initial surprise, this gave a differential light curve for Kepler-13A that displayed a much higher scatter than the raw light curve, about 700\,ppm versus about 450\,ppm, even though the systematic trends seemed to be mostly gone. Upon further consideration, we realized that this was because our raw HST/WFC3 photometry -- unlike ground-based photometry -- was dominated by photon, rather than systematic noise. Thus, while dividing the Kepler-13A by the -13BC light curve removed the systematic trends, it also added the photon noise scatter of the Kepler-13BC in quadrature to the photon noise scatter in Kepler-13A.

We therefore applied a low-pass filter to the Kepler-13BC light curve, to capture the systematic trends in that system's photometry while removing most of the photon noise. Specifically, we used a second-order Butterworth filter with a cutoff at 0.02 days. As shown by the red line in the middle panel of Figure \ref{detrendplot}, the thus filtered Kepler-13BC light curve appeared to correctly capture the trends that we were interested in. We then generated our differential light curve for the white light Kepler-13A data by dividing the raw Kepler-13A photometry by this filtered -13BC light curve. As shown in the bottom panel of Figure \ref{detrendplot}, this removed most of the systematics and made the broadband eclipse clearly visible during both visits.

Another advantage of using Kepler-13BC to make a differential light curve for Kepler-13A is that we were able to make effective use of the first orbit of observations in both of our visits. Typically the entire first orbit of a visit is discarded for precise eclipse or transit observations, due to the increased ramp and thermal breathing affects. In our case, it is a relatively simple matter to remove most of these affects using our differential comparison. The use of the first orbit's data was particularly helpful in the April visit, as without the first orbit we would not have had any pre-eclipse baseline observations.

\begin{deluxetable}{lcl}
\tablecaption{Prior Values for Kepler-13Ab's Properties from \cite{shporer2014}}
\tablehead{\colhead{Parameter} & \colhead{Units} & \colhead{Value}}
\startdata
$T_S$\tablenotemark{a}\dotfill &Predicted $T_S$ (\bjdtdb)\dotfill & $2456776.23411\pm0.00008$\\
$P$\dotfill &Orbital period (days)\dotfill & $1.76358799\pm3.7\times10^{-7}$\\
$e\cos{\omega}$\tablenotemark{b}\dotfill & \dotfill & $-0.00015\pm0.00004$\\
$e\sin{\omega}$\tablenotemark{b}\dotfill & \dotfill & $0.0\pm0.00005$\\
$\cos{i}$\dotfill &Cosine of inclination\dotfill & $0.0714\pm0.008$\\
$R_{P}/R_{*}$\dotfill &Radius ratio\dotfill & $0.0845\pm0.0012$\\
$a/R_{*}$\dotfill &Scaled semimajor axis\dotfill & $4.4\pm0.16$\\
\hline
$M_{P}$\tablenotemark{c}\dotfill &Planet mass (\mj)\dotfill & $6.52\pm1.58$\\
$R_{P}$\tablenotemark{c}\dotfill &Planet radius (\rj)\dotfill & $1.406\pm0.038$
\enddata
\tablenotetext{a}{For the April visit. During fitting we calculate the eclipse time for the October visit using this $T_S$ and advancing by 95 times the orbital period, as described in Section 3.2.}
\tablenotetext{b}{Estimated using the eclipse time and duration given in \cite{shporer2014} and the first order approximations for $e\cos{\omega}$ and $e\sin{\omega}$ given in \cite{winn2010}.}
\tablenotetext{c}{Not a fitting parameter, but provided for reference.}
\label{tab:broadbandpriors}
\end{deluxetable}

For the spectrally-resolved data we also created differential light curves -- but instead of differencing against Kepler-13BC, we differenced against Kepler-13A itself. We were motivated to do so out of a concern that the different placement of the two stars on the detector, plus their slightly different intrinsic spectra, would generate spurious spectral signatures if we differenced against Kepler-13BC. We applied the same low-pass filter as we used in the broadband data to the broadband Kepler-13A photometry, and divided each of our spectrally-resolved light curves by this filtered broadband photometry. 

In our spectrally-resolved data we therefore did not measure the absolute eclipse of Kepler-13Ab, but rather the change in eclipse depth as a function of wavelength.

In both the broadband and spectrally-resolved differential light curves there was residual systematic noise. We chose to fit these residual correlations using a Gaussian Process (GP) regression model, as the traditional parametric methods for dealing with HST/WFC3 systematics were not applicable to our differential light curves.

A GP regression models the observed data as random draws from a multivariate Gaussian distribution about some mean function. As a result, GPs are able to directly model the possible covariances between data points by populating the non-diagonal elements of the covariance matrix which defines the multivariate Gaussian distribution. This is in contrast, for example, to a $\chi^2$ fitting process, which models the data as random draws from a univariate Gaussian distribution and assumes no covariance between data points --- aside from what is inserted via detrending functions. For more detail, \cite{rasmussen2006} provide a thorough mathematical overview of GP methods. \cite{gibson2012} introduced them in the context of astronomical time series observations using archival NICMOS observations of HD 189733 b, though they have a longer history in the general astronomical community \cite[e.g.,][]{way2006}. Recently, \cite{cartier2017} used a GP regression to model HST/WFC3 eclipse observations of WASP-103b, and we follow a similar approach.

\begin{figure}[t]
\vskip 0.00in
\includegraphics[width=1.1\linewidth,clip]{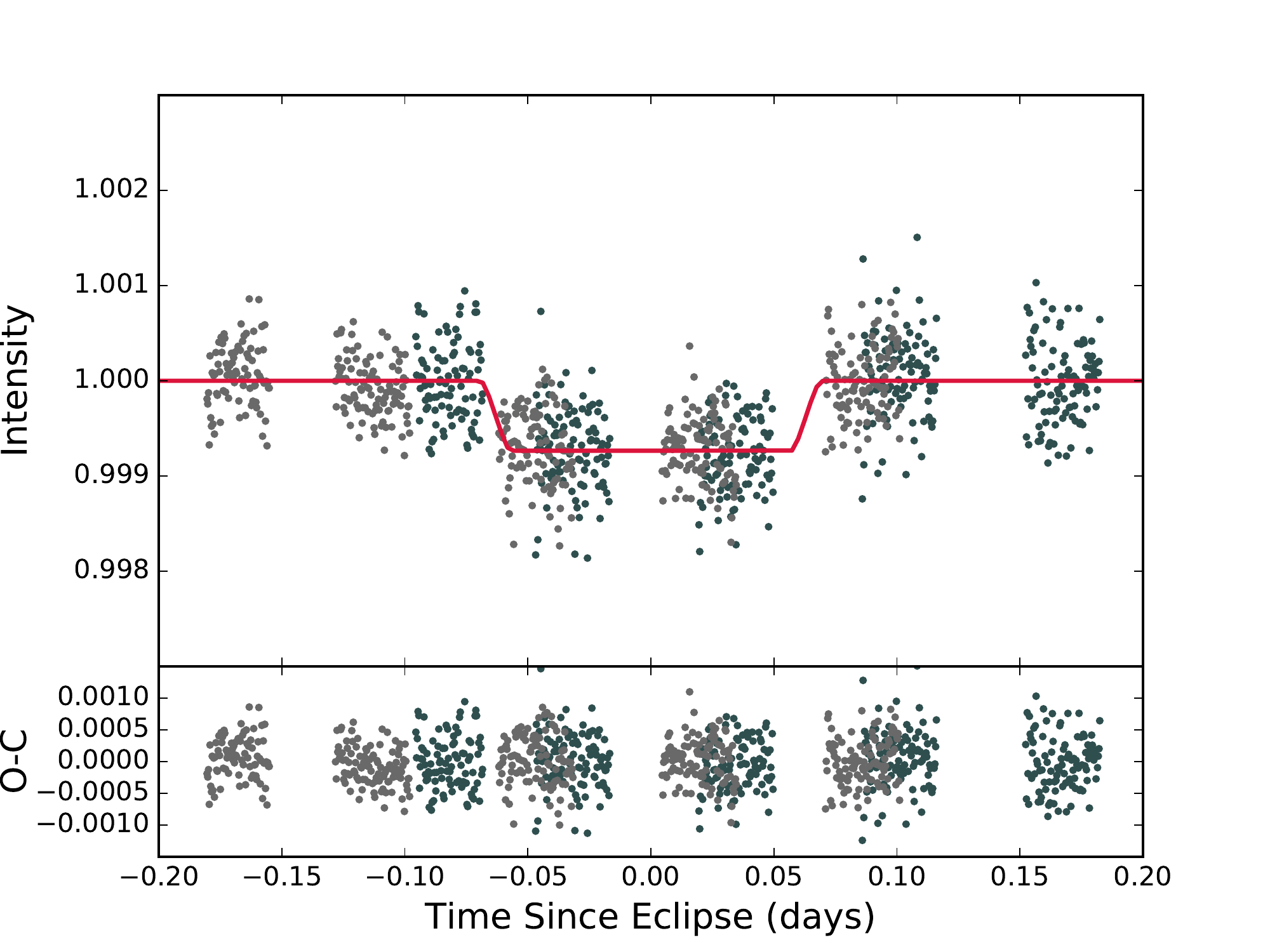}
\vskip -0.0in
\caption{The best fit eclipse model for the combined broadband data from the April (grey-green) and October (grey) visits gives an eclipse depth of $\delta=734\pm28$\,ppm. This depth uncertainty is approximately 1.25 times what one expects from photon noise statistics.}
\vskip -0.05in
\label{niceplot}
\end{figure}

We defined our GP model using the notation from \cite{gibson2012}. For each visit, we have a vector of $N$ observed fluxes, $f = (f_1,..., f_N)$, and times, $t = (t_1,..., t_N)$. Additionally, we recorded $K$ state parameters at each time $t$ with the state vector $x = (x_{t,1},...,x_{t,K})^T$. We combined these state parameter vectors for each of our $N$ observations in the $N \times K$ matrix, $X$. The multivariate Gaussian distribution underlying our GP model was defined by a combination of a mean function, which in our case is the physical eclipse model $E(t,\phi)$, and a covariance matrix $\Sigma(X,\theta)$. We used $\phi$ to denote the set of physical parameters describing the eclipse, and $\theta$ for the set of ``hyperparameters'' used to generate the covariance matrix from the $X$ state parameters. The joint probability distribution of our observed data $f$ was then
\begin{equation}\label{eq:3110}
p(f|X,\theta,\phi)=\mathcal{N}[E(t,\phi),\Sigma(X,\theta)],
\end{equation}
where $\mathcal{N}$ represents the multivariate Gaussian distribution. Our GP model thus depended upon on an eclipse model $E(t,\phi)$, and a generating function -- referred to as the covariance kernel -- for the covariance matrix $\Sigma(X,\theta)$. We generated our GP covariance matrices and calculated the GP likelihoods using the George python package \citep{george2014}.

While we used the same eclipse model to fit both the broadband and spectrally-resolved datasets, we chose slightly different covariance kernels for each set of observations.

\subsubsection{The Eclipse Model, Parameters, and Priors}

We used a \cite{mandel2002} eclipse model as the mean function in our GP regression; specifically the implementation in the \textsc{BATMAN} python package \citep{kreidberg2015}. For our broadband data, we fit for the time of the secondary eclipse ($T_S$), the orbital period (as $\log[P]$), $e\cos{\omega}$, $e\sin{\omega}$, the cosine of the orbital inclination ($\cos i$), the radius of the planet in stellar radii ($R_P/R_*$), the semi-major axis in units of the stellar radii (as $\log[a/R_*]$), and the depth of the secondary eclipse ($\delta$). This gave our broadband eclipse model eight parameters:
\begin{equation}\label{eq:31110}
\phi_\mathrm{white} = (T_S,\log P,e\cos\omega,e\sin\omega,\cos i,R_P/R_*,\log a/R_*,\delta).
\end{equation}

\begin{figure}[t]
\vskip 0.00in
\includegraphics[width=1.1\linewidth,clip]{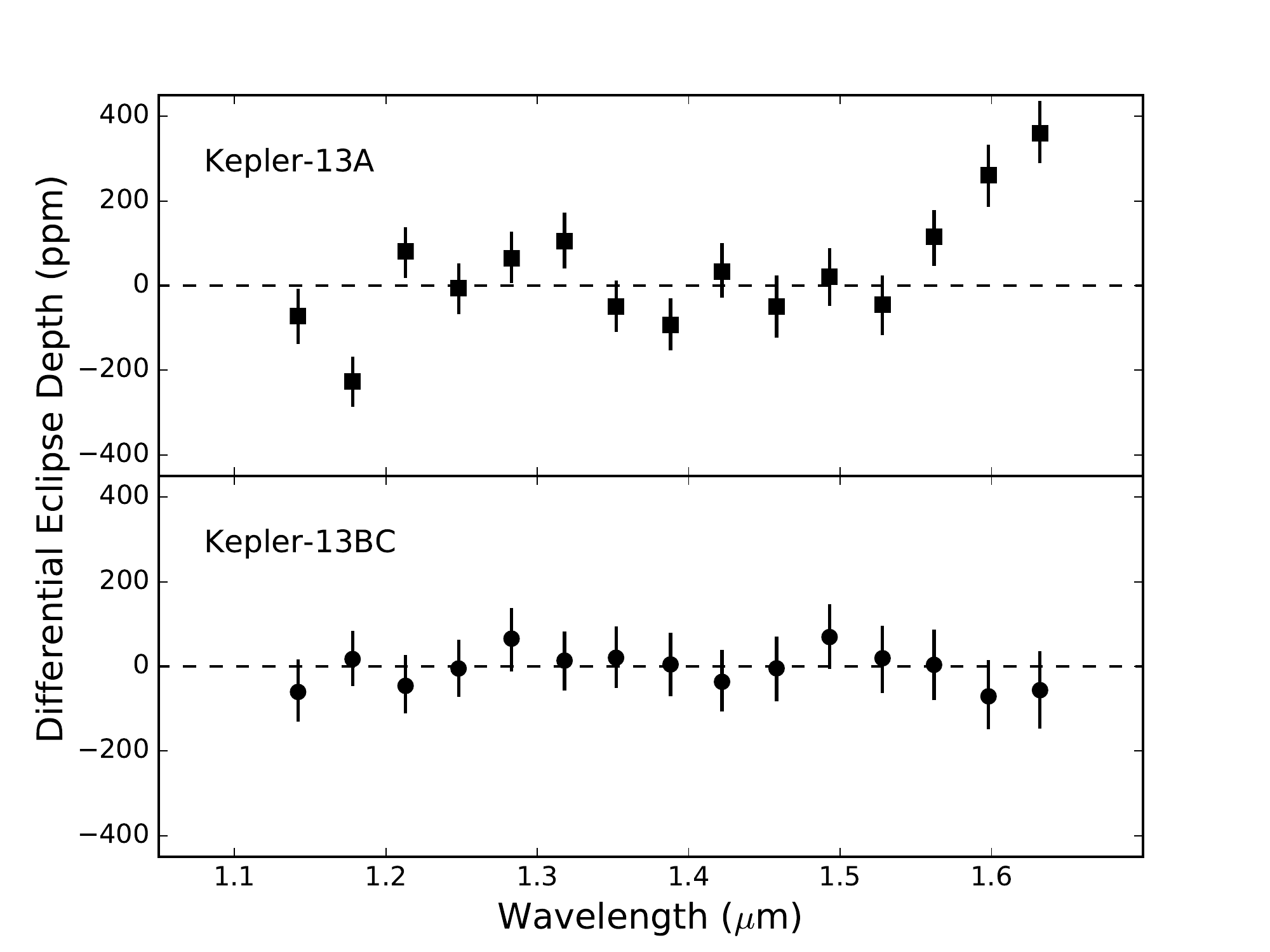}
\vskip -0.0in
\caption{The differential eclipse depths for Kepler-13A (top panel) show variation with wavelength, with $\Delta\chi^2/\mathrm{dof}=4.6$ with 14 degrees of freedom. This corresponds to a $5.9\sigma$ detection of variation. As a check, we also attempt to measure variation in Kepler-13BC using the same fitting method. The differential depths for Kepler-13BC are consistent with zero, with $\Delta\chi^2/\mathrm{dof}=0.3$ with 14 degrees of freedom. Note that both of these spectra are differential measurements against the absolute broadband depth displayed by Kepler-13A ($\delta=734\pm28$\,ppm) and Kepler-13BC ($\delta=-11\pm32$\,ppm).}
\vskip -0.05in
\label{diffplot}
\end{figure}

Based on \cite{shporer2014}'s measurements of the system, we have strong prior expectations for all of these parameters, except for the eclipse depth $\delta$ (Table \ref{tab:broadbandpriors}). In our fitting, we implement each of these as Gaussian priors on these seven parameters. We do not impose any prior on the eclipse depth, as we lack any observations of Kepler-13Ab's eclipse at these wavelengths, which implicitly imposes a uniform prior.

\begin{figure*}[t]
\vskip 0.00in
\includegraphics[width=1.05\linewidth,clip]{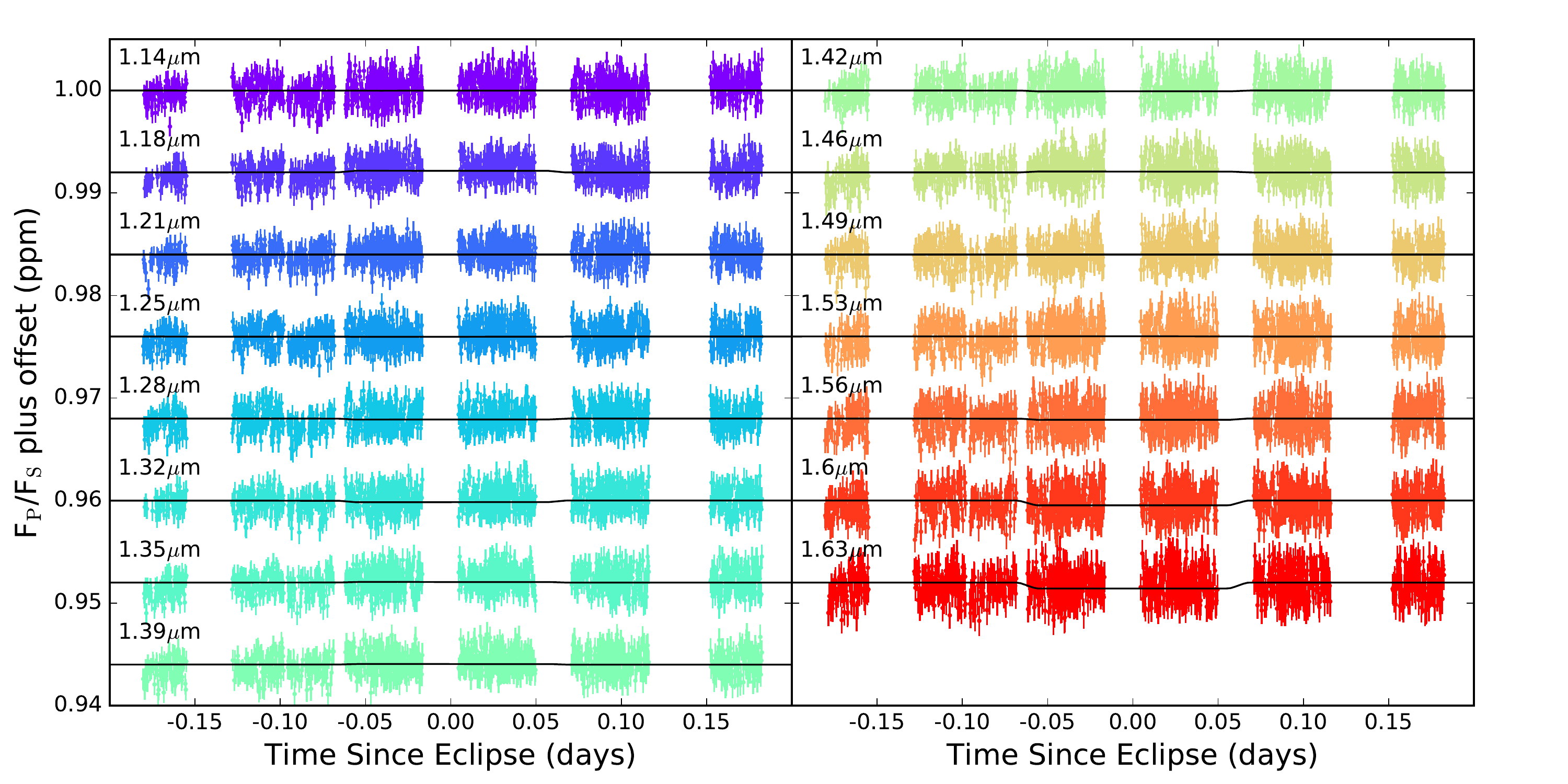}
\vskip -0.0in
\caption{The best fit differential eclipses to our spectrally resolved data. Since we are measuring the eclipse depth relative to the absolute broadband eclipse depth, all of the depths here are relatively shallow, and many are negative.}
\label{stackedplot}
\end{figure*}

For the spectrally-resolved observations the implementation of our eclipse model remains the same, but we only fit for the time of the secondary eclipse and the eclipse depth. All of the other parameters we fix to the values we determine from the broadband fit. This gives our spectrally resolved eclipse model only two parameters:
\begin{equation}\label{eq:31120}
\phi_\mathrm{spec} = (T_S,\delta).
\end{equation}
Again, we impose no prior on the eclipse depths, and we use the measured time of eclipse and its associated uncertainty from the broadband fit as a prior on $T_S$ for the spectrally-resolved fits.  

\subsubsection{GP Covariance Kernel and Hyperparameter Priors}

To model our broadband observations we used a linear combination of a squared-exponential kernel and a periodic kernel -- both as a function of time. A squared-exponential kernel is usually regarded as a generic choice for generating a GP covariance matrix, as it leads to smooth variations as a function of the generating variable. Our intent was to use it to model the residual background temporal trends. We added on a periodic kernel, which is effectively a squared-exponential of a sine function, to model the repeatable covariances between points within an individual orbit.

The point-wise covariances between the observations at times $t_i$ and $t_j$, which collectively made up our broadband $N \times N$ covariance matrix $\Sigma_{\mathrm{white}}$ were then
\begin{equation}\label{eq:31210}
\Sigma_{\mathrm{white}}(t,\theta_{\mathrm{white}}) = \Sigma_{\mathrm{SqExp}}(t,\theta) + \Sigma_{\mathrm{Per}}(t,\theta),
\end{equation}
where
\begin{equation}\label{eq:31220}
\Sigma_{\mathrm{SqExp}}(t,\theta_{\mathrm{white}}) = A_t\, \exp\left[-\frac{(t_i-t_j)^2}{L_t^2}\right],
\end{equation}
and
\begin{equation}\label{eq:31230}
\Sigma_{\mathrm{Per}}(t,\theta_{\mathrm{white}}) = A_p\, \exp\left[\frac{\sin^2(\pi[t_i-t_j]/p)}{L_p^2} \right].
\end{equation}
We used $\theta_{\mathrm{white}}=\{A_t,L_t,A_p,L_p,p\}$ to denote the hyperparameters used to compute the covariances. These are covariance amplitudes, $A_t$ and $A_p$, the covariance length scales $L_t$ and $L_p$. We set the period of the periodic kernel to be $p\equiv95.664$\,minutes. This was the measured orbital period of HST on the dates of our observations, based on archival two-line-elements provided to us by the United States Strategic Command's Joint Space Operations Center.

For our spectrally resolved data, we used a slightly simplified covariance kernel with only a periodic component,
\begin{equation}\label{eq:31240}
\Sigma_{\mathrm{spec}}(t,\theta_{\mathrm{spec}}) = A_p\, \exp\left[\frac{\sin^2(\pi[t_i-t_j]/p)}{L_p^2} \right],
\end{equation}
where we have $\theta_{\mathrm{spec}}=\{A_p,L_p,p\}$ hyperparameters for the spectrally-resolved data, with the same definitions as in the broadband case. 

\begin{figure*}[t]
\vskip 0.00in
\includegraphics[width=1.0\linewidth,clip]{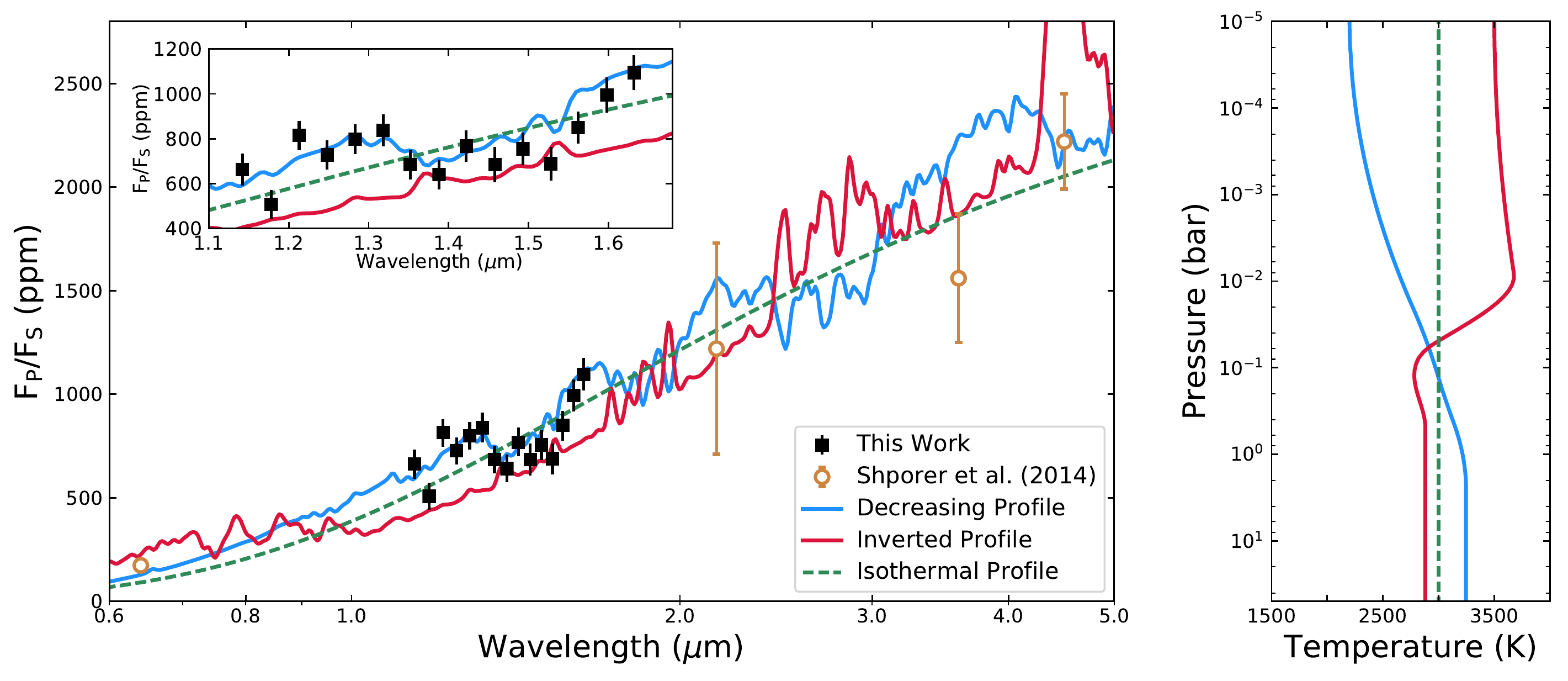}
\vskip -0.0in
\caption{Our modeling of Kepler-13Ab's eclipse depths finds that the observations are well reproduced by an atmospheric model with a monotonically decreasing temperature profile with $\chi^2=29.31$ ($\chi^2/\mathrm{dof}=1.72$, or $2.15\,\sigma$). We exclude the best-fit isothermal model atmosphere, with a temperature of 3000\,K, at $3.84\,\sigma$, and we exclude an inverted temperature profile at $10.4\,\sigma$. The two non-isothermal models assume Solar molecular abundances, but differ in having either Solar TiO/VO (inverted), or zero TiO/VO (decreasing).}
\label{modelplot}
\end{figure*}

We imposed no priors on the members of either set of hyperparameters, other than to require that both $L_t$ and $L_p$ be greater than the eclipse's ingress and egress time of 0.01279 days. This ensured that the GP regression model did not treat the eclipse itself as correlated noise.

\subsection{Fitting Process and Results}

To determine the best fits to the broadband and spectrally-resolved data we maximized a log-likelihood function that consisted of the GP model likelihood and the an additional term based on the priors in Table \ref{tab:broadbandpriors}. For the GP model the log-likelihood for a given set of parameters $\phi$ and $\theta$ was
\begin{equation}\label{eq:3210}
\log p_\mathrm{GP}(r|X,\theta,\phi) = -\frac{1}{2}\bm{r}^T\Sigma^{-1}\bm{r} - \frac{1}{2}\log|\Sigma| - \frac{N}{2}\log(2\pi),
\end{equation}
where $\bm{r}=f - E(t,\phi)$ is the vector of the residuals of our observed data ($f$) from our eclipse model ($E$) defined in Section 3.1.1. This log-likelihood follows directly from our definition of the GP model as a multivariate Gaussian in Equation (\ref{eq:3110}).

All of the physical parameters for which we applied a prior based on previous observations of the system we used Gaussian priors, and we added the log-likelihood of these priors to our GP log-likelihood to compute the total log-likelihood for a given model as
\begin{equation}\label{eq:3220}
\log p_\mathrm{tot}(\phi,\theta|f,X) = \log p_\mathrm{GP}(r|X,\theta,\phi) + \sum \log p_\mathrm{prior}. 
\end{equation}

Our fitting process was composed of two stages for both the broadband and the spectrally-resolved data: an initial Nelder-Mead maximization of Equation (\ref{eq:3220}), followed by an MCMC exploration of the likelihood space around this maximum to determine parameter uncertainties and to verify that we had identified the global likelihood maximum. To conduct the MCMC fitting we used the \textsc{emcee} Python package \citep{dfm2013}. Our MCMC runs consisted of a 500 step burn-in, followed by a 5000 step production run using 30 walkers. We initialized the walkers by scattering them about the initial Nelder-Mead maximum using random draws from Gaussian distributions in each parameter with $1\sigma$ widths equal to corresponding prior widths. At the end of the production runs we calculated the Gelman-Rubin statistic and judged the MCMC to have converged if the statistics for all the parameters were less than 1.1.

We fit the two visits simultaneously using the same set of physical parameters and hyperparameters. We experimented with using differing hyperparameters for each of the visits, but we found that this did not change the results above our final $1\sigma$ uncertainties, nor did it substantially alter the uncertainties themselves. For the spectrally-resolved data we did allow for different hyperparameters between different spectral channels.

Tables \ref{tab:broadbandresults} and \ref{tab:spectralresults} list the results for the broadband eclipse, and the spectrally-resolved eclipses, respectively. Our best fit to the broadband data gave a broadband eclipse depth of $\delta=734\pm27$\,ppm (Figure \ref{niceplot}). The standard deviation in the residuals to the best fit model was 436\,ppm, which was nearly equal to the median per-point flux uncertainty of 400\,ppm from photon statistics. The uncertainty on the broadband measurement was 1.5 times what one would expect based on pure photon noise alone.

The differential depths we measured in our spectrally-resolved light curves (Figure \ref{diffplot} and Figure \ref{stackedplot}) had a median uncertainty of 64\,ppm, which is 1.2 times the photon noise expectation. The median standard deviation of the residuals to the best fit models for the spectrally-resolved data was approximately 1250\,ppm. The differential eclipse depths showed statistically significant variation away from a flat line, with $\Delta\chi^2=64.3$, or $\Delta\chi^2/\mathrm{dof}=4.6$ with 14 degrees of freedom. This corresponds to a $5.6\sigma$ detection of variation with wavelength.

\begin{figure*}[t]
\vskip 0.00in
\includegraphics[width=1.0\linewidth,clip]{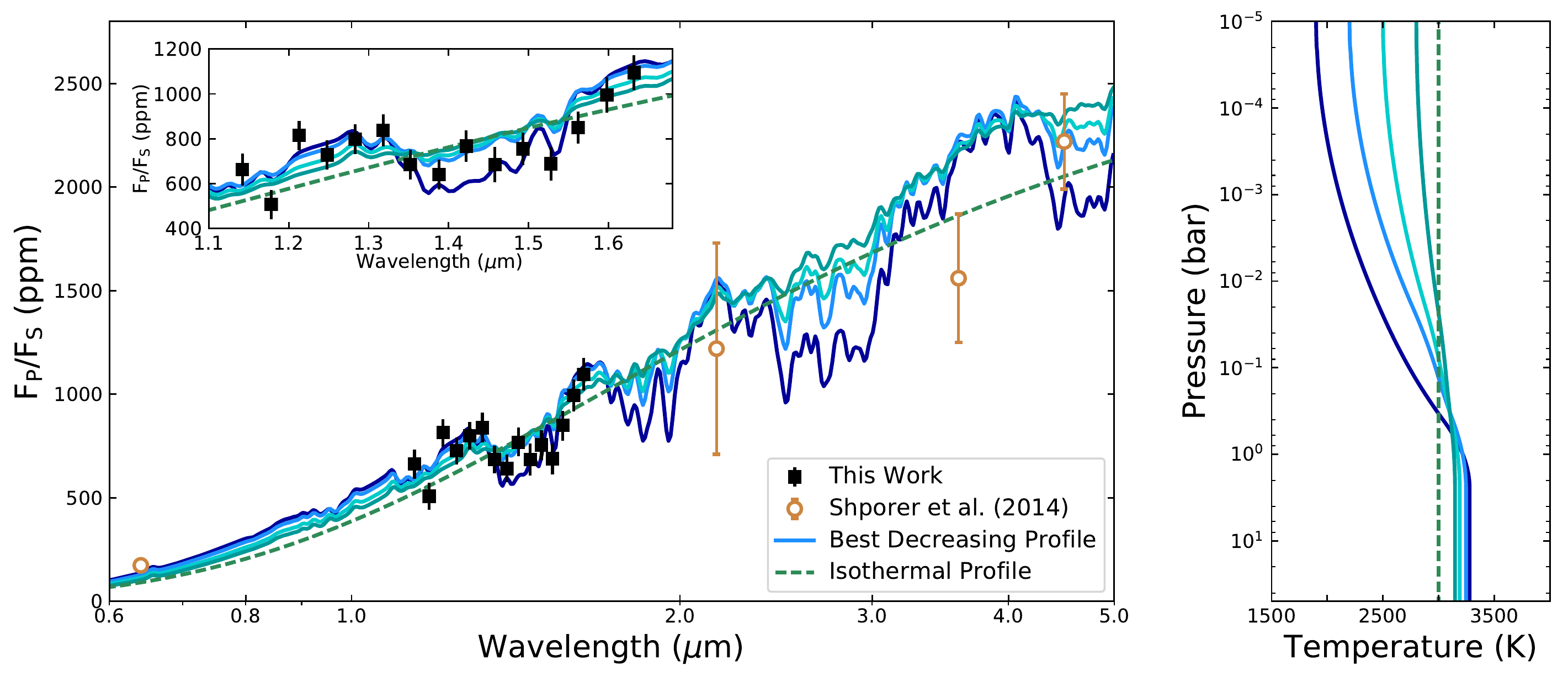}
\vskip -0.0in
\caption{Out of a range of monotonically decreasing temperature profiles, our best fit profile (light blue) is consistent with the observations at $2.14 \,\sigma$. The profile with a stronger temperature slope (dark blue) is also consistent, at $2.24 \,\sigma$. The green-blue temperature profile closest to the purely isothermal profile is only marginally consistent with the observations at $2.96\,\sigma$.}
\label{multiplot}
\end{figure*}

As a verification of our results we also performed the exact same fitting procedures, using the exact same priors, on broadband and spectrally-resolved light curves from Kepler-13BC. As expected, the broadband Kepler-13BC light curve gave an eclipse depth of $\delta_{\mathrm{BC}}=-10\pm 34$\,ppm. As shown in Figure \ref{diffplot}, the differential spectrally-resolved eclipse depths for Kepler-13BC are also all consistent with zero, with $\Delta\chi^2/\mathrm{dof}=0.3$ with 14 degrees of freedom. 

\section{Modeling and Analysis}

We considered the results of our eclipse observations from two perspectives. First, we modeled our measured eclipse depths and those of \cite{shporer2014} using exoplanet atmosphere models. Second we compared the dayside emission spectrum of Kepler-13Ab to isolated field brown dwarfs, since the relatively high mass of Kepler-13Ab (6.5\mj) and the young age of the system (about 500 Myr), means that the planet should have some residual luminosity.

\subsection{Atmospheric Modeling}

We modeled the measured eclipse depth of Kepler-13Ab using the hot Jupiter atmospheric model described in \cite{madhu2009} and \cite{madhu2012}. This model is comprised of a \textsc{1D} plane parallel atmosphere in hydrostatic equilibrium and local thermodynamic equilibrium (LTE). The emergent spectrum is computed using a \textsc{1D} line-by-line radiative transfer solver in the planetary atmosphere. The atmospheric pressure-temperature (PT) profile and chemical composition are free parameters of the model, with 6 parameters for the PT profile and a parameter each for each chemical species included in the model. The range of molecules considered and the sources of opacity are discussed in \cite{madhu2012}. For these models we set the mixing ratios of the major molecular constituents ($\mathrm{H}_2\mathrm{O}$, $\mathrm{CH}_4$, $\mathrm{CO}_2$, CO) to the chemical equilibrium Solar values. The inverted temperature profile in Figure \ref{modelplot} used Solar TiO and VO abundances, while the decreasing temperature profile set the TiO and Vo abundances to zero.

By way of comparison, the model spectra we use have the same qualitative features as the \cite{burrows2008} models shown in Figure 11 of \cite{shporer2014}. The primary modeling difference between our results and those of \cite{shporer2014} is that our WFC3 measurements give a higher average dayside temperature, leading to higher temperatures in our model atmospheres. Compared to the \cite{fortney2008} models from \cite{shporer2014}, our WFC3 spectrum is best fit by the ``no TiO'' case, but we find a much more pronounced water feature at 1.4$\mu$m than predicted by the Fortney models.

We find that our observations are relatively well reproduced by an atmospheric model with a monotonically decreasing temperature profile (Figure \ref{modelplot}), with $\chi^2=29.31$ ($\chi^2/\mathrm{dof}=1.72$, or $2.15\,\sigma$). We exclude the best-fit isothermal model atmosphere, with a temperature of 3000\,K, with $\chi^2=46.98$ ($\chi^2/\mathrm{dof}=2.76$, or $3.84\,\sigma$), and we exclude an inverted temperature profile $\chi^2=133.45$ ($\chi^2/\mathrm{dof}=7.85$, or $10.9\,\sigma$). We also tested a range of monotonically decreasing temperature profiles that stepped towards and away from a purely isothermal profile (Figure \ref{multiplot}). The green-blue profile closest to isothermal in Figure \ref{multiplot} is only marginally consistent with the observations ($2.96\,\sigma$), and is the upper limit allowed by the data.

It is important to note that our WFC3 observations and \cite{shporer2014}'s observations only probe pressure levels in the planetary atmosphere between about $\sim$\,$10^{-2}$ and $\sim$\,$10^0$\,bar. Based solely on these observations, it is therefore possible that a temperature inversion could exist on Kepler-13Ab's dayside higher up and at lower pressure levels.

\subsection{Brown Dwarf Spectral Type Matching}

Due to the relatively high mass of Kepler-13Ab, we were interested to compare its dayside emission spectrum to a brown dwarf of similar effective temperature. Besides investigating the general spectral differences between the dayside of a hot Jupiter and a brown dwarf, we also wished to see if the measured surface gravity of Kepler-13Ab coincided with the surface gravity one would estimate using spectral matching to brown dwarf templates, or using spectral indicators for surface gravity.

Compared to its isolated brown dwarf equivalent the dayside of Kepler-13Ab is, needless to say, considerably hotter. From its age and mass we would expect that Kepler-13Ab would have a surface effective temperature of approximately 400\, to 800\,K if it did not receive any insolation from its host star \citep{burrows2003}. But due to the propinquity of Kepler-13A, the dayside of the planet appears to be closer in temperature to a mid to late M-dwarf. We therefore compared our emission spectrum to a series of spectral templates from the SpeX Library covering M4 to T9 spectra for field objects, and a set of low-gravity templates covering M5 to T9 from \cite{allers2013}. To do so, we converted the normalized flux measurements in the templates to a set of simulated eclipse depths by assuming Kepler-13A was a 7650\,K blackbody, and then scaling the resulting relative flux based on the estimated $J$-band surface brightnesses for Kepler-13A and the spectral templates from the BT-Settl models \citep{allard2011} using the \cite{caffau2011} values for Solar abundances. 

\begin{figure*}[t]
\vskip 0.00in
\includegraphics[width=1.0\linewidth,clip]{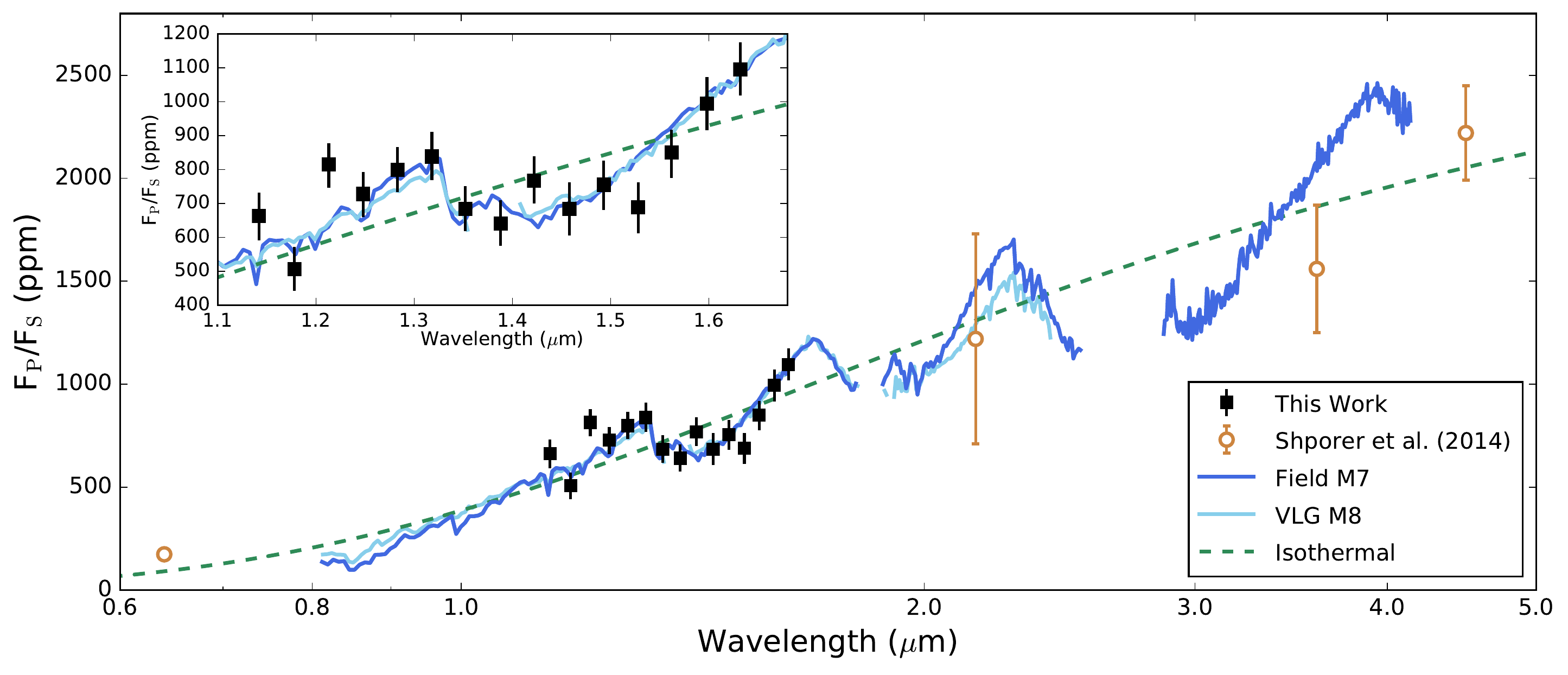}
\vskip -0.0in
\caption{Our WFC3 emission spectrum for Kepler-13Ab (black points), and the eclipse observations from \cite{shporer2014} (gold points) together with the best fit field and low-gravity brown dwarf spectral templates. We find that the dayside of Kepler-13Ab has a spectral type of M7$\pm$2 using both the high-gravity field and low-gravity templates, with an almost identical goodness of fit ($\chi^2/\mathrm{dof}=1.52$ and $\chi^2/\mathrm{dof}=1.71$, respectively). We also show the best fit isothermal model, which has a temperature of 3000\,K.}
\label{bdplot}
\end{figure*}

We find that the dayside of Kepler-13Ab has a rough spectral type of M7$\pm2$ using both the field and low-gravity templates. Using only our fifteen WFC3 measurements, we find $\chi^2=21.19$ ($\chi^2/\mathrm{dof}=1.51$, or $1.66\,\sigma$) for the field template and $\chi^2=20.40$ ($\chi^2/\mathrm{dof}=1.70$, or $1.88\,\sigma$) for the low-gravity template. As shown in Figure \ref{bdplot}, in both cases this result is driven by the presence of the 1.4\um \water absorption feature. We note though, that the apparent improvement on the goodness-of-fit of these template spectra relative to the atmosphere models in Section 4.1 is due to the fact that here we are only using the WFC3 in our comparisons, due to a lack of wavelength coverage in the spectral templates.

The similarity of the higher-gravity field templates and the low-gravity templates in this temperature range means that we are not able to postdict Kepler-13Ab's surface gravity using either template matching or the surface gravity indicators suggested by \cite{allers2013}. Nevertheless, the general agreement of both the field and low-gravity templates indicates that despite the heavy insolation Kepler-13Ab receives, its atmosphere appears similar to isolated brown dwarf atmospheres. Since isolated brown dwarfs are primarily heated from within, and have correspondingly monotonically decreasing atmospheric temperature-pressure profiles, this strengthens our conclusion from the atmospheric models that Kepler-13Ab's dayside is neither isothermal, nor does it possess a stratospheric temperature inversion at any pressure level.

\section{Discussion}

\subsection{Timing of the Eclipse}

\cite{shporer2014} measured a center time for Kepler-13Ab's eclipse that was very close to being half of an orbital period away from the transit center time, with a displacement of $-2.6\pm7.5$\,s. As they noted, due to the light travel time across the diameter of Kepler-13Ab's orbit, for a perfectly circular orbit one would expect the eclipse time to actually be delayed by $+34\pm0.7$\,s. The apparent earliness of the eclipse implies a slightly eccentric orbit, and is what drives the non-zero value of $e\cos{\omega}$ listed in Table \ref{tab:broadbandpriors}.

As described in Section 3.1.1, we used the predicted eclipse time from \cite{shporer2014} as a prior on our broadband fitting process. Since our observations poorly sample the predicted eclipse ingress and egress (i.e., Figure \ref{niceplot}), this prior on $T_S$ dominates in our fits and we simply recover it in the posterior distribution for $T_S$ from our MCMC fitting. In our adopted fit, we therefore find that the eclipse center time is $-1.8\pm7.5$\,s earlier than the transit center time plus exactly half the orbital period, which as we expected is effectively identical to the value measured by \cite{shporer2014}.

To determine how well our HST/WFC3 can independently constrain the eclipse center time, we conducted an additional broadband fit using no prior on the eclipse center. This gave $T_S=2456776.2310\pm0.0025$, which is $-269\pm216$\,s earlier than the transit center time plus exactly half the orbital period. The large uncertainties on this timing offset make it consistent with the expectations from both \cite{shporer2014} and the perfectly circular case, so our observations are not able to meaningfully determine the eccentricity of Kepler-13Ab's orbit. 

\subsection{The Dayside Albedo of Kepler-13Ab}

One complication in interpreting the optical eclipse depth for Kepler-13Ab is the relatively high amount of reflected light expected from the planet. Specifically, the eclipse depth from reflected light alone should be $\delta_{\mathrm{ref}}=A_g(R_p/a)^2=A_g(316)$\,ppm, where $A_g$ is the geometric albedo of the dayside at a particular wavelength. Since the Kepler-band eclipse measured by \cite{shporer2014} is $173.7\pm1.8$\,ppm, it is possible -- though unlikely -- that the optical eclipse is entirely due to reflected light. For our NIR observations, the theoretical expectation is that hot Jupiters at $\approx3000$\,K should have dayside geometric albedos of effectively zero at these wavelengths \citep{sudarsky2000}. We therefore did not consider possible signatures of reflected light in our WFC3 spectra, and excluded the Kepler eclipse point from our evaluation of our thermal emission models.

Based on those thermal emission models that assume the WFC3 and further NIR data contain no reflected light, we calculate that Kepler-13Ab should have an eclipse 133\,ppm deep in the Kepler bandpass. The difference between this prediction for the thermal emission and the observed depth of $173.7\pm1.8$\,ppm implies that Kepler-13Ab has a geometric albedo of $A_g = 0.12$ in the Kepler bandpass. This is substantially lower than the value found by \cite{shporer2014}, $A_s = 0.33$. The difference occurs because our atmosphere models place the interior isotherm at a higher temperature than in 2750\,K blackbody used by \cite{shporer2014} in their albedo calculation, which causes the planetary thermal emission to be higher in the optical.

Our measured geometric albedo and the expectations from \cite{sudarsky2000} are thus both consistent with our assumption that the WFC3 and other NIR eclipse depths contain no substantial components from reflected light. This is in line with what one would expect from other measurements of hot Jupiters' eclipses, most of which have $0.05<A_s<0.2$ in the Kepler bandpass \citep{heng2013}. Note, though, that \cite{heng2013} had to assume the amount of thermal emission present in the Kepler bandpass for their analysis, and could not measure it as we have done here, which makes this only a general comparison. Additionally, the high stellar insolation that Kepler-13Ab receives makes it possible that Kepler-13Ab may have different reflective properties compared to the cooler planets considered by \cite{heng2013}. One way to test this would be to observe an optical eclipse spectrum for the planet. In the case that Kepler-13Ab has a substantial optical albedo, the optical eclipse spectrum would be a combination of the reflectance spectrum \citep[expected to increase towards the blue, e.g.][]{sudarsky2000}, and the thermal emission spectrum (expected to decrease towards the blue), which would make the observed eclipse spectrum relatively flat.

\subsection{Lack of an Inversion}

As described in the Introduction, the lack of clear stratospheric temperature inversions for planets cooler than approximately 3000\,K led \cite{haynes2015} to recently suggest that that TiO/VO driven inversions may only be present in the atmospheres of extremely hot giant planets. \cite{beatty2016}'s spectrally-resolved observations of the $H$-band eclipse of the transiting brown dwarf KELT-1b (3200\,K) showed a monotonically decreasing PT profile. Due to the high surface gravity of KELT-1b (22 times that of Jupiter), this led them to suggest that surface gravity also plays a role in the presence of a thermal inversion in hot Jupiters. Specifically, they argued that this was evidence for the cold-trap methods of sequestering TiO/VO described by \cite{spiegel2009} and \cite{parmentier2013}.

Briefly, both \cite{spiegel2009} and \cite{parmentier2013} describe a process of TiO/VO gas particles condensing and gravitationally settling out of the upper atmosphere. \cite{spiegel2009} envisioned this as a ``vertical'' cold-trap, where TiO/VO gas on the dayside of a hot Jupiter randomly crosses the condensation boundary in the atmosphere, condenses, and falls into the planetary interior. \cite{parmentier2013} considered a ``day-night'' cold-trap, where TiO/VO molecules condense on the cooler nightside of a hot Jupiter, and also settle into the interior. In both cases the efficiency of a cold-trap is determined by the interplay of the rate of gravitational settling and the strength of some vertical lofting mechanism to bring TiO/VO condensates back into the upper atmosphere where they can re-vaporize. Both analyses found that under reasonable assumptions these two cold-traps should be able to deplete a hot Jupiter's stratosphere of gas-phase TiO/VO.

In the cases of both Kepler-13Ab and KELT-1b, the dayside temperature profiles both remain too hot at depth to allow for a vertical cold-trap to exist. Down to pressures of $10^2$ bar, at no point are either of these profiles expected to cross to cooler than the TiO or VO condensation curves, which vary between approximately 2200\,K at $10^2$ bar down to approximately 1600\,K at $10^{-5}$ bar. The most recent modeling of such a process was done by \cite{parmentier2016}, who found that a vertical cold-trap of TiO/VO was mostly effective in planets with equilibrium temperatures less than 1900\,K, which is several hundred degrees cooler than the planets we consider. It is also interesting to note that simulations by \cite{tremblin2017} show that vertical cold-traps may be impossible in hot Jupiters due to advection in their atmospheres. In either case, if a cold-trap process is occurring in these atmospheres - and in the atmospheres of the other extremely hot giant planets - it is very probably caused by a day-night cold-trap.

To first order, the rate of gravitational settling in a hot Jupiter's atmosphere will be given by the free-fall timescale within that atmosphere. This will be the scale height of the atmosphere divided by the free-fall terminal velocity of condensates,
\begin{equation}\label{eq:4110}
\tau_\mathrm{ff}=\frac{H}{V_{\mathrm{term}}} = \left(\frac{k_{\mathrm{B}}T}{\mu_{\mathrm{m}}g}\right)\left(\frac{2a^2g(\rho_\mathrm{p}-\rho)}{9\eta}\right)^{-1}.
\end{equation}
Here $T$ is the atmospheric temperature, $\mu_m$ is the mean molecular weight of the atmosphere, $g$ is the gravitational acceleration, $a$ is the particle radius, $\rho_\mathrm{p}$ is the particle density, $\rho$ is the atmospheric density, and $\eta$ is the viscosity of the gas. If we make the assumption that mean molecular weight and viscosity of hot Jupiters' atmospheres, as well as the condensate and atmospheric density, are all effectively the same, then the free-fall timescale goes as
\begin{equation}\label{eq:4120}
\tau_\mathrm{ff} \propto Ta^{-2}g^{-2}.
\end{equation}

Typically, turbulent diffusion and large scale vertical mixing are treated as the dominant vertical lofting mechanisms for condensates in the upper portion of a hot Jupiter's atmosphere, but the exact efficiency of these processes is poorly understood. By analogy to molecular diffusion, the efficiency of vertical mixing is parameterized by the effective diffusion coefficient, $k_{zz}$ \citep{banks1973}. Unlike the molecular diffusion coefficient, however, the value of $k_{zz}$ is the result of inherently chaotic processes that are difficult to model. While mixing length theory can be used to derive analytic estimates for $k_{zz}$ \citep[e.g.,][]{kzzmixing1985} in the convective regions of a hot Jupiter's atmosphere, these estimates are not applicable to the the upper, radiative, portion of the atmosphere probed by eclipse observations. This make the precise value of $k_{zz}$ difficult to predict, and typical values of $k_{zz}$ used in brown dwarf modeling can cover three to four orders of magnitude \citep[e.g.,][]{saumon2007}.  

If we make the assumption that the efficiency of vertical mixing is approximately the same in all hot Jupiters' atmospheres, then for fixed values of $a$, the condensate size, cold-traps will therefore be less efficient in hotter atmospheres with higher values of $T$, but dramatically more efficient as surface gravity increases. 

\begin{figure}[t]
\vskip 0.00in
\includegraphics[width=1.05\linewidth,clip]{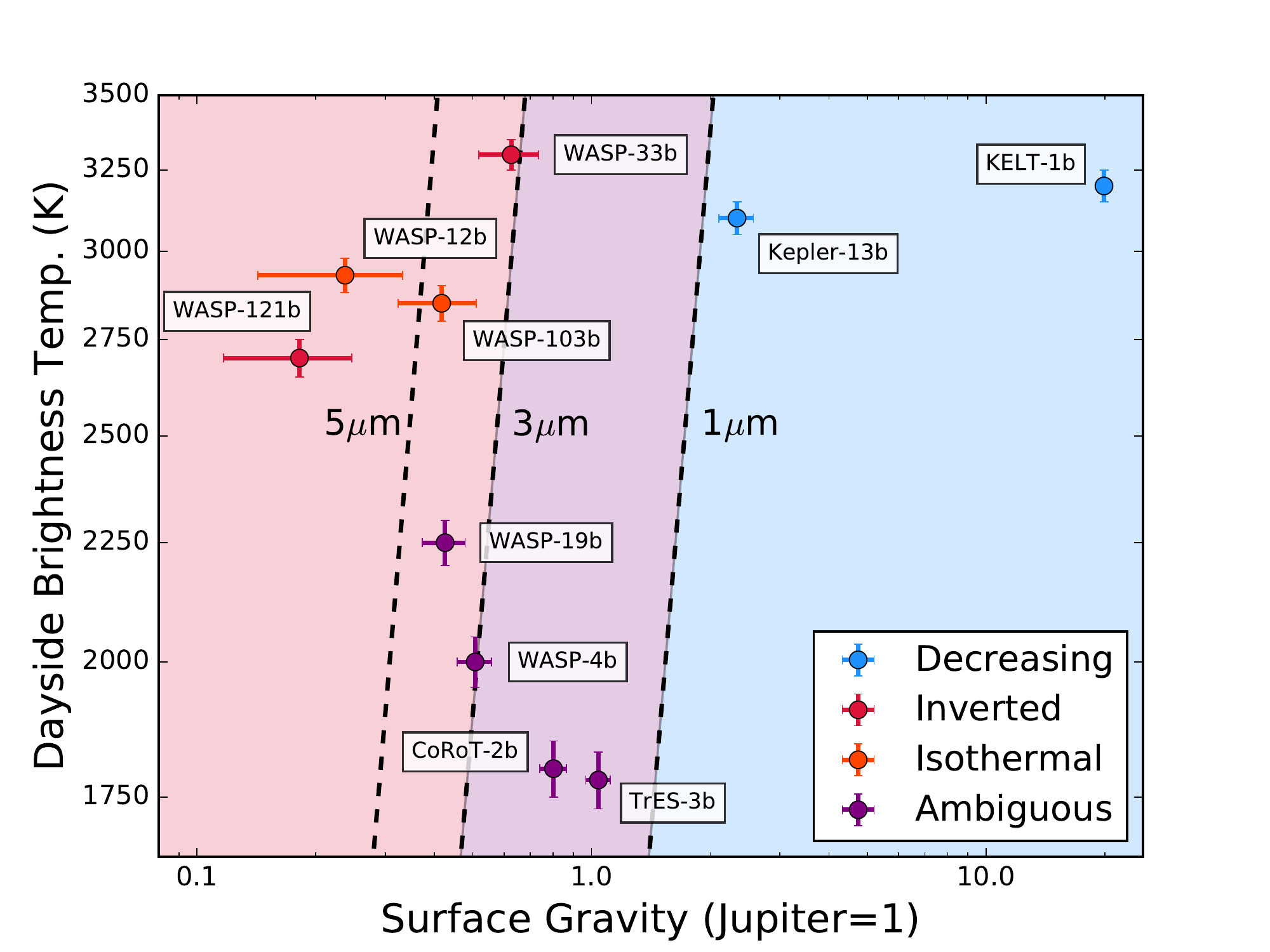}
\vskip -0.0in
\caption{The dashed lines indicate the approximate locii where cold-trap processes should clear TiO/VO from a hot Jupiter's atmosphere and inhibit a stratospheric temperature inversion for different mean condensate sizes, based on the scaling of Equation (\ref{eq:4120}) from the results of \cite{parmentier2013} on HD 189733b. Based on existing observations, the red shaded region roughly corresponds to parameter space we would expect to see inverted or isothermal atmospheres, while the blue shaded region should have decreasing PT profiles. The middle purple region is ambiguous, owing to the range of allowable particle sizes. Note that it is entirely possible that the mean condensate size changes over this parameter space, or varies individually by planet.}
\vskip -0.1in
\label{logglogt}
\end{figure}

Though more detailed modeling is necessary to precisely assess the role of cold-trap processes in Kepler-13Ab's atmosphere, let us proceed from the conclusion in \cite{parmentier2013} that for particle sizes greater than ``a few microns'' a day-night cold trap is capable of depleting gas-phase TiO/VO in HD 189733b's atmosphere. Using the scaling in Equation (\ref{eq:4120}, we may extrapolate from this point to other temperatures and surface gravities, and compare against the other hot Jupiters with spectrally-resolved NIR emission measurements.

Based on their 1.1\um to 1.7\um emission and the modeling in their respective papers, we categorized the other planets with spectrally-resolved NIR eclipse observations as having an inverted PT profile (WASP-33b, WASP-121b), an isothermal profile (WASP-12b, WASP-103b), or as being ambiguous (WASP-19b \citep{bean2013}, WASP-4b \citep{ranjan2014}, TrES-3b \citep{ranjan2014}, and CoRoT-2b \citep{wilkins2014}). Of these four ambiguous planets, TrES-3b and CoRoT-2b both show WFC3 spectra consistent with both an isothermal and decreasing PT profile. The temperatures for the isothermal models of both (1800\,K and 1780\,K, respectively) are very close to the condensation temperature of TiO/VO. Depending upon the exact temperature structure of their atmospheres, it is therefore possible that TiO and VO have condensed out of their upper atmospheres independently of any possible cold-trap mechanisms. For WASP-4b, the WFC3 measured spectrum agrees with both a 2000\,K isothermal model and a carbon-rich, monotonically decreasing, model. WASP-19b was observed from the ground by \cite{bean2013}, and their emission spectrum is not precise enough to distinguish the temperature structure of the planet. We categorized Kepler-13Ab as having a monotonically decreasing PT profile, as well as KELT-1b. Extrapolating from HD 189733 b according to Equation (\ref{eq:4120}), we then plotted these eight planets alongside the limits of where we would expect cold-traps to inhibit inversions for particle sizes of $a=1\mu$m, $a=3\mu$m, and $a=5\mu$m. 

As shown in Figure \ref{logglogt}, this set of classifications and predictions relatively well represents the planets with HST/WFC3 observations of their dayside emission. In the shaded red region, where we would expect the atmospheres to be inverted or isothermal for reasonable particle sizes, are WASP-33b, WASP-12b, and WASP-103b. Similarly, Both Kepler-13Ab and KELT-1b lie in the region where we would expect to see decreasing PT profiles assuming the condensate particles grow larger than $a=1\mu$m.

We note that \cite{heng2013} have also considered the efficiency of cold-traps, but instead choose to compare vertical terminal velocities to vertical mixing velocities to determine if a condensate sinks into the planetary interior, without considering the atmospheric scale height. Based on numerical simulations of the vertical mixing velocity in planets cooler than 1750\,K, this leads \cite{heng2013} to estimate that the efficiency of cold-trap processes should be roughly independent of temperature, and should scale inversely with condensate size and surface gravity. This nicely contrasts with our Equations \ref{eq:4110} and \ref{eq:4120}, which propose a different scaling. Which of these is correct - if any - would therefore provide us with good insight into the dominant processes with hot Jupiters' atmospheres, and could indicate whether vertical mixing velocities are roughly constant, as we have assumed, or if they vary coherently with planetary properties, as assumed by \cite{heng2013}.

Currently, the global maximum size for condensate particles in the atmospheres of extremely hot Jupiters with daysides near 3000\,K is poorly constrained by atmospheric models. Based on the Rayleigh scattering signatures seen in transmission spectra the maximum particle size at high altitudes along the planetary limb appears to be on the order of 0.1$\mu$m \citep{wakeford2015}. But 3D models of HD 187933 b's ($T_{eq}\approx1500$\,K) atmosphere by \cite{lee2016} show a wide range of possible particle sizes as a function of longitude, latitude, and depth within the atmosphere. \cite{parmentier2016} recently analyzed Kepler-band albedo estimates from eclipse measurements made by \cite{esteves2015}, in an effort to constrain condensate properties on planetary daysides and at higher pressures than those probed by transmission observations. \cite{parmentier2016} found that an average condensate size of 0.1$\mu$m was able to replicate the reflective properties of the cooler planets ($T_{eq}\approx1600$\,K), but that the two hotter planets in the \cite{esteves2015} data set near  $T_{eq}\approx2200$\,K required a larger average condensate size. What the condensate size is in a hot Jupiter with a dayside near 3000\,K has not been well modeled.

Based on the toy model diagrammed in Figure (\ref{logglogt}), with more spectrally-resolved observations of hot Jupiters' daysides, it may be possible to observationally constrain the general condensate size in these planets atmospheres by mapping out where stratospheric temperature inversion occur in extremely hot Jupiters. Though this assumes that cold-traps are the dominant mechanism to inhibit inversion, it will also be necessary to determine if one can even speak of single ``condensate size'' for all hot Jupiters. To more adequately assess the role that cold-traps play in hot Jupiters' atmospheres, we therefore need more two- or three-dimensional atmosphere models, such as those in \cite{parmentier2016} and \cite{lee2016}, that include the variation of temperature structure as a function of longitude and latitude, models of particle and condensate growth processes within hot Jupiters' atmospheres, and we need spectrally-resolved observations of more systems to validate the results of both.  

\section{Summary and Conclusions}

We observed two secondary eclipses of the transiting hot Jupiter Kepler-13Ab using HST/WFC3 on UT 2014 April 28 and UT 2014 October 13. We were able to separate the two primary components of the stellar system, Kepler-13A and Kepler-13BC, in our staring-mode grism spectroscopy. Using \textsc{wayne} simulations of both components, we extracted dilution-corrected spectrophotometry for each component by subtracting off the simulations from our observed images. 

The presence of Kepler-13BC allowed us to use its broadband light curve as a comparison star, and the resulting differential light light curve of the Kepler-13A system showed very little of the systematic trends typically associated with HST/WFC3 spectrophotometry (Figure \ref{detrendplot}. Together with a Gaussian Process regression model, this allowed us to measure the broadband eclipse depth as $\delta=734\pm27$\,ppm (Table \ref{tab:broadbandresults}). This corresponds to an average dayside brightness temperature of 3000\,K. As a check, we ran the same extraction and fitting process on Kepler-13BC and measured an eclipse depth of $-10\pm34$\,ppm, which is consistent with zero, as expected.

For the spectrally-resolved eclipse depths we made another differential measurement -- but this time against the broadband light curve of Kepler-13A itself. We made this choice so that our results were not influenced by possible differences in the spectral responses of Kepler-13A and Kepler-13BC. As a result, our spectrally-resolved fits using another Gaussian Process regression model measure the differential eclipse depth of Kepler-13Ab as a function of wavelength. We find that the planetary emission spectrum displays significant ($5.6\sigma$) variation with wavelength (Table \ref{tab:spectralresults}). We again performed the same extraction and fitting on the corresponding spectrally-resolved data of Kepler-13BC, and measure differential eclipse depths consistent with zero, again as we expected (Figure \ref{diffplot}).

Our eclipse spectrum shows the 1.4\um \water feature in absorption, and modeling of this feature and the previous eclipse measurements made by \cite{shporer2014} indicate that the dayside of Kepler-13Ab possesses a monotonically decreasing temperature-pressure profile at the pressure levels observed ($\sim$\,$10^{-2}$ to $\sim$\,$10^0$\,bar). It is possible that Kepler-13-Ab's dayside temperature profile does become inverted at lowerpressure levels than probed by the observations, but our inference that Kepler-13Ab's dayside temperature is monotonically decreasing is further supported by the fact that the shape and amplitude of our HST/WFC3 spectrum shows that the dayside of Kepler-13Ab appears similar to the spectrum of a field M7 dwarf.

We contend that the dual facts that Kepler-13Ab possesses a decreasing temperature-pressure profile and a relatively high surface gravity support the hypothesis of \cite{beatty2016} that both surface gravity and temperature play a role in determining the presence of a stratospheric temperature inversion in hot Jupiters. Specifically, in high surface gravity planets such as Kepler-13Ab, the characteristic free-fall time within the atmosphere is substantially shorter (Equation \ref{eq:4120}). This should, in turn, substantially increase the efficiency of a day-night \citep{parmentier2013} cold-trap process, thereby sequestering the TiO/VO molecules available to cause an inversion in the interior of the planet. 

Of the nine hot Jupiters with HST/WFC3 eclipse observations and dayside brightness temperatures hotter than 1750\,K, the scaling of where an inversion should occur implied by Equation \ref{eq:4120} and a reasonable range of condensate particle sizes seems to reproduce the observations fairly well (Figure \ref{logglogt}). If we accept the assumption that cold-trap processes are the dominant inhibitor of stratospheric temperature inversions in these giant planets, then more spectrally-resolved observations of planets within this parameter space could allow us to observationally constrain the maximum condensate size possible in a hot Jupiter's dayside, atmosphere. This would provide us with constraints on the cloud growth processes and the bulk vertical mixing of the atmosphere.

To properly assess this, however, we will also need to work on two- and three-dimensional atmosphere models. The current one-dimensional models are generally not capable of treating the local atmospheric variations that are presumable important in setting the average condensate size in an atmosphere. Furthermore, while developing higher dimensional models would have an immediate impact here, we also note that it would also allow us to better treat the observed temperature gradients across planetary daysides and from day to night. Pushing for higher dimensional models, and more detailed, spectrally-resolved, observations of hot Jupiters' atmospheres, will give us a much better window into the vertical temperature structure of these planets.

\acknowledgements

We would like to thank the anonymous referee for their detailed response, which improved the quality of the paper.

These observations were made as a part of GO Program 13308 with the NASA/ESA Hubble Space Telescope at the Space Telescope Science Institute, which is operated by the Association of  Universities for Research in Astronomy, Inc., for NASA, under the contract NAS 5-26555.

T.G.B., M.Z., and J.T.W. were partially supported by funding from the Center for Exoplanets and Habitable Worlds. The Center for Exoplanets and Habitable Worlds is supported by the Pennsylvania State University, the Eberly College of Science, and the Pennsylvania Space Grant Consortium. A. T. was supported by the ERC project ExoLights (617119). M.Z. and J.T.W. acknowledge NASA Origins of Solar Systems grant NNX14AD22G.

This work has made use of NASA's Astrophysics Data System, the Exoplanet Orbit Database and the Exoplanet Data Explorer at exoplanets.org \citep{exoplanetsorg}, the Extrasolar Planet Encyclopedia at exoplanet.eu \citep{exoplanetseu}, the SIMBAD database operated at CDS, Strasbourg, France \citep{simbad}, and the VizieR catalog access tool, CDS, Strasbourg, France \citep{vizier}.

\begin{deluxetable*}{lcl}
\tablecaption{Median Values and 68\% Confidence Intervals for the Broadband Eclipse Observations}
\tablehead{\colhead{~~~Parameter} & \colhead{Units} & \colhead{Value}}
\startdata
\sidehead{GP Hyperparameters:}
                ~~~$A_t$\dotfill &Sqr. Exp. covariance amplitude\dotfill & $2.4\times10^{-6}\,_{-1.6\times10^{-6}}^{+6.5\times10^{-5}}$\\                
             ~~~$L_t$\dotfill &Sqr. Exp. covariance length-scale\dotfill & $1.48_{-0.32}^{+0.92}$\\
             	  ~~~$A_p$\dotfill &Periodic covariance amplitude\dotfill & $1.4\times10^{-7}\,_{-6.7\times10^{-8}}^{+1.8\times10^{-7}}$\\
              ~~~$L_p$\dotfill &Periodic covariance length-scale\dotfill & $2.38_{-0.57}^{+0.69}$\\                  
\sidehead{Eclipse Model Parameters:}
                        ~~~$T_S$\dotfill &Eclipse Time (\bjdtdb)\dotfill & $2456776.234119_{-0.000079}^{+0.000073}$\\
                 ~~~$\log(P)$\dotfill &Log orbital period (days)\dotfill & $0.24639714\pm9.2\times10^{-8}$\\
                                   ~~~$e\cos{\omega}$\dotfill & \dotfill & $-0.00015\pm0.00004$\\
                                   ~~~$e\sin{\omega}$\dotfill & \dotfill & $0.0\pm0.00005$\\                     
                    ~~~$\cos{i}$\dotfill & Cosine of inclination\dotfill & $0.069\pm0.008$\\
     ~~~$R_{P}/R_{*}$\dotfill &Radius of planet in stellar radii\dotfill & $0.0844\pm0.0012$\\
~~~$\log(a/R_{*})$\dotfill &Log semi-major axis in stellar radii\dotfill & $0.6325\pm0.0075$\\
                        ~~~$\delta$\dotfill &Eclipse depth (ppm)\dotfill & $734\pm28$\\
\sidehead{Derived Parameters:}
                           ~~~$P$\dotfill &Orbital period (days)\dotfill & $1.76358803\pm0.00000037$\\
          ~~~$a/R_{*}$\dotfill &Semi-major axis in stellar radii\dotfill & $4.29\pm0.08$\\
                           ~~~$i$\dotfill &Inclination (degrees)\dotfill & $86.04\pm0.44$\\
                                ~~~$b$\dotfill &Impact Parameter\dotfill & $0.296\pm0.031$\\
                     ~~~$T_{FWHM}$\dotfill &FWHM duration (days)\dotfill & $0.1263_{-0.0022}^{+0.0025}$\\
               ~~~$\tau$\dotfill &Ingress/egress duration (days)\dotfill & $0.0119\pm0.0003$\\
                      ~~~$T_{14}$\dotfill &Total duration (days)\dotfill & $0.1382_{-0.0024}^{+0.0027}$\\\
                            ~~~$e$\dotfill &Orbital Eccentricity\dotfill & $0.00016\pm0.00004$\\
           ~~~$\omega$\dotfill &Argument of periastron (degrees)\dotfill & $128_{-300}^{+44}$
\enddata
\label{tab:broadbandresults}
\end{deluxetable*}
\begin{deluxetable*}{cccccc}
\tablecaption{Median Values and 68\% Confidence Intervals for Spectrally-resolved Eclipse Observations}
\tablehead{\colhead{~~~$\lambda$ ($\mu$m)} & \colhead{Abs. Depth (ppm)} & Diff. Depth (ppm) & \colhead{$T_S$ (\bjdtdb-2456770)} & \colhead{$A_p$} & \colhead{$L_p$}}
\startdata
$1.142$ & $662\pm71$ & $-71\pm65$  & $6.234116\pm0.000078$ & $0.000003_{-0.000002}^{+0.000007}$    & $0.12\pm0.075$         \\
$1.178$ & $507\pm65$ & $-227\pm59$ & $6.234122\pm0.000075$ & $0.000004_{-0.000003}^{+0.000016}$    & $0.15\pm0.075$         \\
$1.213$ & $814\pm65$ & $81\pm60$   & $6.234116\pm0.000077$ & $0.000003_{-0.0000003}^{+0.0000008}$  & $0.10_{-0.06}^{+0.08}$ \\
$1.248$ & $728\pm66$ & $-5\pm60$   & $6.234123\pm0.000078$ & $0.0000004_{-0.0000003}^{+0.000001}$  & $0.13\pm0.07$          \\
$1.283$ & $798\pm66$ & $64\pm61$   & $6.234119\pm0.000079$ & $0.000002_{-0.000001}^{+0.000002}$    & $0.10\pm0.065$         \\
$1.318$ & $838\pm70$ & $-104\pm66$ & $6.234113\pm0.000075$ & $0.000001_{-0.0000006}^{+0.000001}$   & $0.11\pm0.070$         \\
$1.352$ & $684\pm67$ & $-50\pm61$  & $6.234118\pm0.000072$ & $0.0000002_{-0.0000001}^{+0.0000006}$ & $0.11\pm0.071$         \\
$1.388$ & $640\pm66$ & $-93\pm61$  & $6.234118\pm0.000073$ & $0.0000002_{-0.0000001}^{+0.0000004}$ & $0.10\pm0.065$         \\
$1.422$ & $767\pm71$ & $33\pm65$  & $6.234119\pm0.000074$ & $0.0000006_{-0.0000005}^{+0.000002}$  & $0.11\pm0.068$         \\
$1.458$ & $684\pm78$ & $-50\pm73$  & $6.234124\pm0.000075$ & $0.0000005_{-0.0000004}^{+0.000001}$  & $0.11_{-0.06}^{+0.08}$ \\
$1.493$ & $755\pm72$ & $21\pm67$   & $6.234131\pm0.000075$ & $0.000001_{-0.0000008}^{+0.000003}$   & $0.13\pm0.072$         \\
$1.528$ & $688\pm75$ & $-46\pm70$  & $6.234123\pm0.000076$ & $0.0000002_{-0.0000001}^{+0.0000004}$ & $0.10_{-0.055}^{+0.075}$\\
$1.562$ & $850\pm71$ & $116\pm66$  & $6.234124\pm0.000075$ & $0.0000004_{-0.0000003}^{+0.000003}$  & $0.09\pm0.007$         \\
$1.598$ & $994\pm78$ & $261\pm73$   & $6.234108\pm0.000078$ & $0.0000003_{-0.0000002}^{+0.000002}$  & $0.12\pm0.07$          \\
$1.632$ & $1094\pm78$ & $291\pm73$   & $6.234119\pm0.000076$ & $0.0000001_{-0.0000001}^{+0.000006}$  & $0.08_{-0.05}^{+0.08}$
\enddata
\label{tab:spectralresults}
\end{deluxetable*}

\end{document}